\documentclass[aps,prl,reprint,letterpaper,showpacs,twocolumn,citeautoscript]{revtex4-1}
\usepackage[english]{babel}
\usepackage{graphicx}
\usepackage[pdftex                     
           ,pagebackref=false          
           ,colorlinks=false           
           ]{hyperref}
\hypersetup{linkcolor=blue,            
            citecolor=red,             
            urlcolor=black}            

\usepackage{amsmath}
\usepackage{mathtools}
\usepackage{eufrak}

\newcommand*{\abinitio}{{\itshape ab initio}}
\newcommand*{\firstprinciples}{first principles}
\newcommand*{\fleur}{\texttt{FLEUR}}
\newcommand*{\wprog}{\textsc{wannier}{\footnotesize{90}}}
\newcommand*{\DM}{Dzyaloshinskii-Moriya}
\newcommand*{\bstruc}{band structure}

\newcommand*{\I}{i}
\newcommand*{\E}[1]{\mathrm{e}^{#1}}
\newcommand*{\D}{\mathrm{d}}

\newcommand*{\vn}{\boldsymbol}

\newcommand*{\gwfshort}{HDWFs}
\newcommand*{\gpos}{\vn \xi}
\newcommand*{\gdlatt}{\vn \Xi}
\newcommand*{\gpara}{\vn \lambda}

\newcommand*{\gwann}{W}
\newcommand*{\gumat}{\mathcal{U}}
\newcommand*{\artpsi}{\zeta}
\newcommand*{\artu}{\rho}
\newcommand*{\uvec}{\hat{\vn e}}

\begin{document}

\begin{abstract}
When using Wannier functions to study the electronic structure of multi-parameter Hamiltonians $H^{(\vn k,\gpara)}$ carrying a dependence on crystal momentum $\vn k$ and an additional periodic parameter $\gpara$, one usually constructs several sets of Wannier functions for a set of values of $\gpara$.
We present the concept of higher dimensional Wannier functions (\gwfshort{}), which provide a minimal and accurate description of the electronic structure of multi-parameter Hamiltonians based on a single set of \gwfshort{}. The obstacle of non-orthogonality of Bloch functions at different $\gpara$ is overcome by introducing an auxiliary real space, which is reciprocal to the parameter $\gpara$. We derive a generalized interpolation scheme and emphasize the essential conceptual
and computational simplifications in using the formalism, for instance, in the evaluation of linear response coefficients. We further implement the necessary machinery to construct \gwfshort{} from \abinitio{} within the full-potential linearized augmented plane-wave method (FLAPW). We apply our
implementation to accurately interpolate the Hamiltonian of a one-dimensional magnetic chain of Mn atoms in two important cases of $\gpara$: (i)~the spin-spiral vector $\vn{q}$, and (ii)~the direction of the ferromagnetic magnetization $\hat{\vn{m}}$. Using the generalized interpolation of the energy, we extract the corresponding values of magneto-crystalline anisotropy energy, Heisenberg exchange constants, and spin stiffness, which compare very well with the values obtained from direct first principles calculations. For toy models we demonstrate that the method of HDWFs can also be used in applications such as the virtual crystal approximation, ferroelectric polarization and spin torques.
\end{abstract}

\setcounter{secnumdepth}{3}
 \title{Higher dimensional Wannier functions of multi-parameter Hamiltonians}
 \author{Jan-Philipp Hanke}
 \email{j.hanke@fz-juelich.de}
 \author{Frank Freimuth}
 \author{Stefan Bl\"ugel}
 \author{Yuriy Mokrousov}
 \date{March 5, 2015}
 \affiliation{Peter Gr\"unberg Institut and Institute for Advanced Simulation,\\Forschungszentrum J\"ulich and JARA, 52425 J\"ulich, Germany}
 \maketitle

\section{Introduction}
\label{sec:introduction}
Maximally localized Wannier functions (MLWFs) have become a widely applied tool in electronic structure calculations \cite{Marzari2012}. Defined as discrete Fourier transformations of Bloch states $\Psi_{\vn k m}$ with respect to crystal momentum $\vn k$, the MLWFs
\begin{equation}
 W_{\vn R n}(\vn r) = \frac{1}{N_{\vn k}}\sum\limits_{\vn k m} \E{-\I \vn k \cdot \vn R}  U_{mn}^{(\vn k)} \Psi_{\vn km}(\vn r)
 \label{eq:WF_def}
\end{equation}
are labeled by the direct lattice vector $\vn R$ and the orbital index $n$. Unitary gauge transformations $U^{(\vn k)}$ as well as the number of $\vn k$-points, $N_{\vn k}$, enter Eq.~\eqref{eq:WF_def}. These orbitals allow for an efficient but remarkably accurate Wannier interpolation of any single-particle operator such as the Hamiltonian $H^{(\vn k)}$. The Wannier interpolation is in particular fruitful in the calculation of linear response coefficients such as the anomalous Hall conductivity, and various Fermi surface properties, which require a fine $\vn k$-mesh for Brillouin zone (BZ) integration \cite{Wang2006,Yates2007,Yao2004}.


It is sometimes necessary to consider a family $H^{(\vn k,\gpara)}$ of Hamiltonians, where $\gpara$ is an additional parameter. In the problem of ferroelectric polarization, for instance, the parameter $\gpara$ indicates relative displacements of the crystal sublattices \cite{King-Smith1993,Vanderbilt1993,Resta1994}. In magnetic systems with non-collinear spin-spiral texture, the additional parameter $\gpara$ can be identified with the spin-spiral vector $\vn q$. The related energy $E(\vn q)$ serves to determine Heisenberg exchange constants \cite{Lezaic2013}. Frequently, the ferromagnetic magnetization direction $\hat{\vn m}$ plays the role of $\gpara$. Such a situation is met in the study of the magneto-crystalline anisotropy energy (MAE), which is the magnitude of the variation of the energy $E(\hat{\vn m})$ with $\hat{\vn m}$. The dependencies of crystal volume, current-induced torques \cite{Freimuth2014,Freimuth2014a,Garello2013,Kurebayashi2014}, and the conductivity tensor on the magnetization direction provide similar examples.

Notably, the anomalous Hall effect (AHE) can exhibit an anisotropy with respect to $\hat{\vn m}$ \cite{Roman2009}. For a given magnetization direction, the corresponding value of the anomalous Hall conductivity is obtained from Wannier interpolation in crystal momentum $\vn k$. The evaluation on a dense $\hat{\vn m}$-mesh requires accordingly the construction of a huge amount of MLWFs -- one set of MLWFs for each $\hat{\vn m}$. Consequently, the accurate calculation of the AHE anisotropy is a rather time-consuming task.

The computation of such linear response quantities would benefit in particular from an interpolation technique based on functions which provide efficient access to the multi-parameter Hamiltonian $H^{(\vn k,\gpara)}$ of the system. For this purpose, the definition of MLWFs, Eq.~\eqref{eq:WF_def}, has to be generalized. We introduce higher dimensional Wannier functions (\gwfshort{}) as Fourier transformations of states $\Phi_{\vn k \gpara m}$ with respect to both crystal momentum $\vn k$ and the additional parameter $\gpara$:
\begin{equation}
\begin{split}
 \gwann_{\vn R \gdlatt n}(\vn r, \gpos) = \frac{1}{N_{\vn k}}\frac{1}{N_{\gpara}} &\sum\limits_{\vn k \gpara m} \E{-\I \vn k\cdot \vn R} \E{-\I \gpara \cdot \gdlatt}\\
&\times \gumat_{mn}^{(\vn k,\gpara)} \Phi_{\vn k \gpara m}(\vn r, \gpos) \, .
\end{split}
\label{eq:gen_WF_def}
\end{equation}
Here, $\gumat^{(\vn k,\gpara)}$ denotes a unitary matrix and the new additional index $\gdlatt$ of the \gwfshort{} is conjugate to $\gpara$ like $\vn R$ is conjugate to $\vn k$. Further, $N_{\gpara}$ is the number of $\gpara$-points and $\gpos$ refers to an auxiliary space variable. In the presence of an additional parameter, the usual Bloch states $\Psi_{\vn k \gpara m}$ are typically not orthogonal, i.e., $\langle \Psi_{\vn k \gpara n} | \Psi_{\vn k^\prime \gpara^\prime m} \rangle \not\propto \delta_{\vn k\vn k^\prime}\delta_{\gpara \gpara^\prime}\delta_{nm}$. To overcome this obstacle and establish the transformation Eq.~\eqref{eq:gen_WF_def}, the introduction of an auxiliary space $\gpos$ is crucial. The auxiliary space is reciprocal to $\gpara$ like real space is reciprocal to crystal momentum $\vn k$. Then, the role of Bloch states is taken in Eq.~\eqref{eq:gen_WF_def} by orthogonal states $\Phi_{\vn k \gpara m}$ in the combined space of $\vn r$ and $\gpos$.

Within an energy window of interest, the family $H^{(\vn k,\gpara)}$ of Hamiltonians can be interpolated using \gwfshort{}. After constructing \gwfshort{} from a coarse $(\vn k,\gpara)$-mesh, we store the information on the multi-parameter Hamiltonian in hopping elements $H_{nm}(\vn R,\gdlatt)$ of \gwfshort{}. Finally, we can obtain $H^{(\vn k,\gpara)}$ on a much denser $(\vn k,\gpara)$-mesh by an inverse Fourier transformation of these hoppings.

Several applications where such an approach would be very fruitful come to mind, of which we mention explicitly the following ones:
(i)~The evaluation of the AHE anisotropy would be simplified by performing a generalized Wannier interpolation for the situation with $\gpara=\hat{\vn m}$; (ii)~Heisenberg exchange constants would be accessible using generalizations of MLWFs in an interpolation of $E(\vn q)$ with respect to the spin-spiral parameter $\gpara=\vn q$; (iii)~Mixed Berry curvatures in real and momentum space have been found to be quantitatively important in materials like MnSi, where they support the formation of non-trivial magnetic textures \cite{Freimuth2013}. An accurate interpolation of the multi-parameter Hamiltonian could be employed in the study of the contributions of mixed Berry curvatures to the Hall effects. Thus, \gwfshort{} would prove useful in the topological characterization of complex magnetic structures; (iv)~Such functions could provide an alternative means of calculating forces in \firstprinciples{} methods where atomic displacements are described by $\gpara$; (v)~Eventually, the framework could allow the treatment of alloys like Fe$_x$Co$_{1-x}$ or even Bi$_x$Sb$_{1-x}$ within the virtual crystal approximation (VCA), with concentration $x$ as parameter $\gpara$.

In this work, we present the formalism of higher dimensional Wannier functions (\gwfshort{}) given by Eq.~\eqref{eq:gen_WF_def}. The problem of non-orthogonality of Bloch states is solved by the introduction of an auxiliary space reciprocal to the $\gpara$-space. Based on \gwfshort{}, we establish a generalized interpolation scheme which provides efficient but accurate access to the multi-parameter Hamiltonian $H^{(\vn k,\gpara)}$ for any desired value of $(\vn k,\gpara)$. The necessary machinery for an \abinitio{} construction of \gwfshort{} is implemented within the FLAPW method to treat consistently multi-parameter Hamiltonians of realistic systems. As proof of principle, we consider the electronic structure of a linear equidistant chain of Mn atoms as a function of (i) the spin-spiral vector $\vn q$, and (ii) as a function of the ferromagnetic magnetization direction 
$\hat{\vn{m}}$. Using the method of HDWFs, we achieve the generalized
interpolation of the first principles Hamiltonian family $H^{(\vn k,\vn q)}$ and $H^{(\vn k,\hat{\vn{m}})}$, which allows for a precise determination of Heisenberg exchange parameters, spin stiffness and magneto-crystalline anisotropy
energy. Within toy models we investigate further promising applications of the formalism such as VCA, current-induced torques, and ferroelectric polarization.

The paper is structured as follows. We begin with a concise review of MLWFs and the Wannier interpolation in Sec.~\ref{sec:std_WFs}. In Sec.~\ref{sec:extend_formalism}, we introduce the formalism of \gwfshort{} and set up the interpolation technique of multi-parameter Hamiltonians. We describe the implementation for constructing \gwfshort{} from \abinitio{} within the FLAPW method in Sec.~\ref{sec:mn_chain}, and present the application of HDWFs to calculating Heisenberg exchange constants and MAE of a
Mn chain in Sec.~\ref{sec:mn_chain} and \ref{sec:mn_chain_mae}, respectively. In Sec.~\ref{sec:model}, we discuss applications of \gwfshort{} for VCA, ferroelectric polarization, and current-induced torques based on toy models. Finally, we conclude this work with a summary.

\section{Review of \texorpdfstring{MLWF\lowercase{s}}{MLWFs}}
\label{sec:std_WFs}
In contrast to the oscillatory and delocalized Bloch states $\Psi_{\vn k n}$, Wannier functions (WFs) provide a more intuitive insight into the nature of crystal bonding \cite{Silvestrelli1998,Posternak2002,Lee2005,Abu-Farsakh2007} and the underlying physical processes due to their real-space localization. The benefit of reformulating the electronic structure problem in terms of WFs is widely exploited in formal developments such as effective model Hamiltonian construction for the study of strongly correlated systems \cite{Anisimov2005,Ren2006,Lechermann2006}. Further, the centers of WFs play a fundamental role in the modern theory of ferroelectric polarization \cite{King-Smith1993,Vanderbilt1993,Resta1994}.

In the definition of MLWFs, Eq.~\eqref{eq:WF_def}, the unitary matrices $U^{(\vn k)}$ are chosen to maximize the real-space localization whereby the resulting orbitals are uniquely determined. One approach to obtain such a gauge thus lies in minimizing the spatial extent $\Omega$ of the WFs:
\begin{equation}
 \Omega = \sum\limits_n \left(\langle W_{\vn 0 n}|r^2|W_{\vn 0 n}\rangle - \langle W_{\vn 0 n}|\vn r|W_{\vn 0 n}\rangle^2\right)\, .
\label{eq:std_WF_spread}
\end{equation}
An algorithm for the spread minimization of WFs was proposed first for the case of isolated groups of energy bands \cite{Marzari1997} but soon generalized to treat entangled bands as well \cite{Souza2001}. The corresponding \wprog{} implementation requires as an input two quantities \cite{Mostofi2008}. First, the overlaps
\begin{equation}
 M_{mn}^{(\vn k,\vn b)}=\langle u_{\vn k m} | u_{\vn k+\vn b \, n}\rangle
\label{eq:overlap_k_std}
\end{equation}
of the periodic parts $u_{\vn k m} = \E{-\I\vn k \cdot \vn r} \Psi_{\vn k m}$ of the Bloch states at neighboring crystal momenta $\vn k$ and $\vn k+\vn b$ have to be provided since they determine centers and spreads of MLWFs. Second, the projections $A_{mn}^{(\vn k)} = \langle \Psi_{\vn k m} | g_n\rangle$ of the Bloch functions onto localized trial orbitals $g_n$ serve as a starting point for the iterative minimization process, which results at the end in MLWFs.

The Wannier interpolation is performed within a certain energy window spanned by the MLWFs \cite{Yates2007}. For this purpose, matrix elements of the single-particle Hamiltonian $H$ between such functions have to be calculated:
\begin{equation}
\begin{split}
H_{nm}(\vn R) &= \langle W_{\vn 0 n} | H | W_{\vn R m} \rangle \\
&= \frac{1}{N_{\vn k}}\sum\limits_{\vn k n^\prime}\E{-\I \vn k \cdot \vn R} \left(U_{n^\prime n}^{(\vn k)}\right)^* \mathcal{E}_{\vn k n^\prime} \,U_{n^\prime m}^{(\vn k)} \, ,
\end{split}
\label{eq:std_WF_hopp}
\end{equation}
where $\mathcal{E}_{\vn k n^\prime}$ stand for the \abinitio{} band energies computed on a coarse $\vn k$-mesh of $N_{\vn k}$ points. Importantly, MLWFs are orthonormal such that $\langle W_{\vn R n} | W_{\vn R^\prime m} \rangle = \delta_{\vn R \vn R^\prime}\delta_{nm}$, which follows from the orthogonality of the Bloch states $\langle \Psi_{\vn k n} | \Psi_{\vn k^\prime m} \rangle = N_{\vn k} \delta_{\vn k \vn k^\prime}\delta_{nm}$ and Eq.~\eqref{eq:WF_def}. Because of the localization of MLWFs, the matrix elements $H_{nm}(\vn R)$ decay rapidly with increasing distance $|\vn R|$. The electronic \bstruc{} can be accessed accurately on a much finer interpolation mesh of $\vn{k}$-points using the hopping elements, Eq.~\eqref{eq:std_WF_hopp}. By an inverse Fourier transformation the interpolated Hamiltonian $H^{(\vn k)}$ is obtained for every desired $\vn k$-point, even if this point is not contained in the coarse mesh of $N_{\vn k}$ points used for constructing MLWFs:
\begin{equation}
 H_{nm}^{(\vn k)} = \sideset{}{'}\sum\limits_{\vn R}\E{\I \vn k\cdot \vn R} H_{nm}(\vn R)\, .
 \label{eq:std_WF_H_int}
\end{equation}
Here, as marked with a dash, the summation is truncated keeping in mind the rapid decay of the hopping elements $H_{nm}(\vn R)$. Eventually, the interpolated Hamiltonian is diagonalized using unitary matrices $V^{(\vn k)}$:
\begin{equation}
 \left[\left(V^{(\vn k)}\right)^\dagger H^{(\vn k)} \, V^{(\vn k)}\right]_{nm} = \mathcal{E}_{\vn k n}\delta_{nm}\, .
\end{equation}
Thus, the Wannier interpolation grants efficient access to the \bstruc{} $\mathcal{E}_{\vn k n}$ for any $\vn k$. Key properties necessary for this interpolation scheme to work are orthonormality as well as real-space localization of the MLWFs.

\section{Extension of the formalism}
\label{sec:extend_formalism}
\subsection{Orthogonality problem}
In the presence of an additional periodic variable $\gpara$, the system under consideration is described by a family of Hamiltonians, where each member $H^{(\gpara)}$ represents the system at a given value of $\gpara$. If we assume that $H^{(\gpara)}$ is lattice periodic at each $\gpara$, the eigenstates of $H^{(\gpara)}$ are Bloch states $\Psi_{\vn k \gpara n}$ carrying a dependence on $\gpara$:
\begin{equation}
 H^{(\gpara)}(\vn r) \Psi_{\vn k \gpara n}(\vn r) = \mathcal{E}_{\vn k \gpara n} \Psi_{\vn k \gpara n} (\vn r) \, ,
\label{eq:h_parameter}
\end{equation}
where $\mathcal{E}_{\vn k \gpara n}$ are the band energies. Since the Hamiltonians $H^{(\gpara)}$ and $H^{(\gpara^\prime)}$ are generally independent, the eigenstates at different
values of $\gpara$ are not necessarily orthogonal, i.e.,
\begin{equation}
 \langle \Psi_{\vn k \gpara n} | \Psi_{\vn k^\prime \gpara^\prime m} \rangle \not\propto \delta_{\vn k \vn k^\prime} \delta_{\gpara \gpara^\prime} \delta_{nm}\, .
\end{equation}
Only at fixed parameter $\gpara$ the orthogonality with respect to crystal momentum is always present such that $\langle \Psi_{\vn k \gpara n} | \Psi_{\vn k^\prime \gpara m}\rangle= N_{\vn k} \delta_{\vn k \vn k^\prime} \delta_{nm}$. As a consequence, discrete Fourier transformations of these Bloch states with respect to $\vn k$ and $\gpara$ do not lead to orthonormal WFs. On the one hand, nonorthogonal WFs can be defined \cite{Skylaris2002} and can even be advantageous due to a stronger real-space localization \cite{He2001}. On the other hand, in our case already the eigenstates are nonorthogonal for $\gpara\neq\gpara^\prime$, leading to additional complications. In particular, when trying to generalize Eq.~\eqref{eq:std_WF_hopp} for the case of these nonorthogonal WFs, we formally encounter matrix elements $\langle \Psi_{\vn k \gpara n} | H | \Psi_{\vn k \gpara^\prime m}\rangle$ the handling of which is not obvious for $\gpara\neq\gpara^\prime$.

\subsection{Solution to the orthogonality problem}
\label{subsec:solution_orthogonality}
\subsubsection{Introduction of an auxiliary space}
To obtain well-localized orthonormal \gwfshort{}, we introduce an auxiliary space $\gpos$ as the reciprocal of the $\gpara$-space. Instead of taking the Bloch states $\Psi_{\vn k \gpara n}(\vn r)$ in the construction of \gwfshort{}, we consider orthogonal states $\Phi_{\vn k \gpara n}(\vn r, \gpos)$ in the composite space $(\vn r, \gpos)$. We define such states as the products of the physical Bloch states and auxiliary orbitals $\artpsi_{\gpara}(\gpos)$:
\begin{equation}
 \Phi_{\vn k \gpara n}(\vn r,\gpos) = \Psi_{\vn k \gpara n}(\vn r) \artpsi_{\gpara}(\gpos) \, .
\label{eq:gen_WF_prod}
\end{equation}
The crucial orthogonality of the product states $\Phi_{\vn k \gpara n}$ is enforced by choosing $\langle \artpsi_{\gpara} | \artpsi_{\gpara^\prime}\rangle = N_{\gpara}\delta_{\gpara\gpara^\prime}$.

\subsubsection{Choice of the auxiliary orbital}
\begin{figure}
 \centering
 \includegraphics{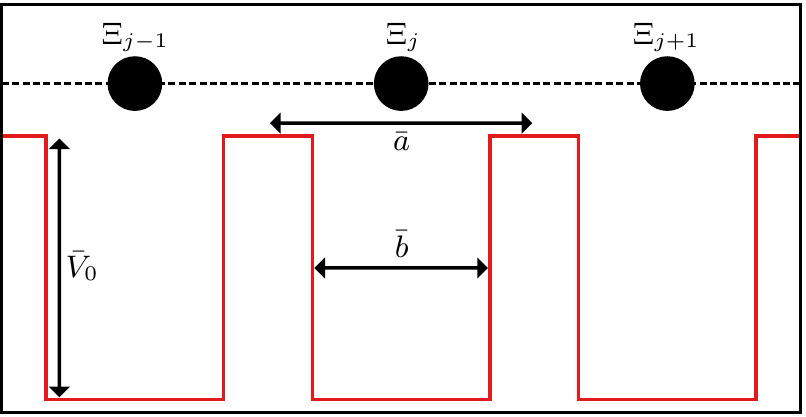}
 \caption{Scalar potential landscape (red solid line) of the one-dimensional lattice defined as series of finite potential wells of depth $\bar V_0$. The well width is $\bar b$, and $\bar a$ stands for the lattice constant. A dashed line indicates the $\xi$ axis.}
 \label{fig:potential_well}
\end{figure}

When constructing the auxiliary orbital, we consider a translationally invariant potential in the auxiliary space as schematically shown in Fig.~\ref{fig:potential_well}. The auxiliary orbital $\artpsi_{\gpara}(\gpos)$ is chosen to be the lowest energy eigenstate of an according lattice periodic Hamiltonian $\bar H$:
\begin{equation}
\bar H(\gpos) \artpsi_{\gpara}(\gpos) = \bar{\mathcal{E}}_{\gpara}\artpsi_{\gpara}(\gpos) \, ,
\label{eq:art_SG}
\end{equation}
where $\bar{\mathcal{E}}_{\gpara}$ represents the lowest energy band in $\gpara$-space associated with the Hamiltonian $\bar H$. The regular lattice in the auxiliary space is modeled using a series of potential wells of depth $\bar V_0$ as depicted in Fig.~\ref{fig:potential_well}. Because the extension to higher dimensions is straightforward, we only discuss the one-dimensional case described by the single-particle Hamiltonian
\begin{equation}
 \bar H(\xi) = -\frac{\hbar^2}{2m}\frac{\D^2}{\D \xi^2} -\bar V_0 \sum\limits_{j} \Theta_{\Xi_j}^{\bar b}(\xi) \, ,
 \label{eq:art_H}
\end{equation}
where $m$ is the electron mass and $\hbar$ is Planck's constant. To simplify notation, we introduced the function
\begin{equation}
\Theta_{\Xi_j}^{\bar b}(\xi) = \Theta\left(\xi-\Xi_j+\bar b/2\right) - \Theta\left(\xi-\Xi_j-\bar b/2\right)\, ,
\label{eq:well_function}
\end{equation}
which cuts out the well region of width $\bar b$ centered around the position $\Xi_j$ with the Heaviside step function $\Theta(\xi)$. Here, the coordinate $\Xi_j = j \bar a$ is defined by the lattice constant $\bar a$ measured along the $\xi$ axis and an integer $j$.

The most convenient and natural choice of the auxiliary orbital is that of a Bloch wave:
\begin{equation}
 \artpsi_{\gpara}(\gpos) = \E{\I \gpara \cdot \gpos} \artu_{\gpara}(\gpos) \, ,
\label{eq:auxiliary_state}
\end{equation}
where $\artu_{\gpara}(\gpos)$ is a $\gpos$-periodic function normalized to the unit cell in the auxiliary space: $\langle \artu_{\gpara} | \artu_{\gpara^\prime}\rangle = \delta_{\gpara\gpara^\prime}$. It follows that the auxiliary orbitals are orthogonal, i.e., $\langle \artpsi_{\gpara} | \artpsi_{\gpara^\prime}\rangle = N_{\gpara} \delta_{\gpara\gpara^\prime}$, where the integration is performed in a supercell of $N_{\gpara}$ unit cells in the auxiliary space.

We can solve the Schr\"odinger equation to the one-dimensional Hamiltonian Eq.~\eqref{eq:art_H} for $\artpsi_{\lambda}$ numerically using a plane-wave basis, with the potential depth $\bar V_0$ chosen to strongly suppress the tunneling between different wells. Alternatively, we can also arrive analytically at the expression for the lowest energy eigenstate to Eq.~\eqref{eq:art_H} in the deep-well limit $\bar V_0\rightarrow \infty$ by starting from a single-well solution:
\begin{equation}
 w(\xi) = \begin{dcases}
              \sqrt{\frac{2}{\bar b}}\,\cos \frac{\pi\xi}{\bar b} & \text{, if } |\xi|<\bar b/2 \\
              0 & \text{, else}
             \end{dcases} \, .
\label{eq:single_well_solution}
\end{equation}
The auxiliary orbital is found as an inverse Fourier transformation of the Wannier-like function $w$ with respect to the positions $\Xi_j$. In accordance with the Bloch theorem, Eq.~\eqref{eq:auxiliary_state}, the lattice periodic part $\rho_\lambda$ assumes the form
\begin{equation}
 \rho_\lambda(\xi) = \E{-\I \lambda\xi} \sum\limits_{j} \E{\I \lambda \Xi_j} w(\xi-\Xi_j) \, .
\label{eq:auxiliary_solution}
\end{equation}
Eventually, overlaps between periodic parts $\rho_\lambda$ to different parameter values $\lambda$ and $\lambda+\tau$ are important ingredients for constructing \gwfshort{}. Such overlaps read
\begin{equation}
 \langle \rho_\lambda | \rho_{\lambda+\tau}\rangle = \frac{8\pi^2\sin\left(\tau \bar b/2\right)}{4\pi^2\tau \bar b - \tau^3 \bar b^3} = 1+\frac{\tau^2 \bar b^2}{24\pi^2}(6-\pi^2)+\mathcal{O}(\tau^4)\, ,
 \label{eq:auxiliary_overlap}
\end{equation}
where $\tau$ plays a similar role like $\vn b$ in Eq.~\eqref{eq:overlap_k_std} and the $\xi$ integration is performed in one unit cell in auxiliary space. From the above Taylor expansion it follows that $\langle \rho_\lambda | \partial_\lambda \rho_\lambda\rangle =0$ and $\langle \rho_\lambda | \partial_\lambda^2 \rho_\lambda\rangle = (6-\pi^2)\bar b^2/(12\pi^2)$.

\subsubsection{Product states and composite Hamiltonian}
Essentially, the definition of \gwfshort{}, Eq.~\eqref{eq:gen_WF_def}, is based on the products $\Phi_{\vn k \gpara n}(\vn r, \gpos)$ of Bloch states and the auxiliary orbitals discussed above. Exploiting Eq.~\eqref{eq:auxiliary_state}, we can rewrite the product states of Eq.~\eqref{eq:gen_WF_prod} as
\begin{equation}
 \Phi_{\vn k \gpara n}(\vn r, \gpos) = \E{\I \vn k \cdot \vn r} \E{\I \gpara \cdot \gpos} \varphi_{\vn k \gpara n}(\vn r,\gpos) \, ,
\label{eq:product_states_end}
\end{equation}
where 
\begin{equation}
\varphi_{\vn k \gpara n}(\vn r,\gpos) = u_{\vn k \gpara n}(\vn r) \artu_{\gpara}(\gpos)
\label{eq:product_periodic_end} 
\end{equation}
are lattice periodic. Such product states are orthogonal also in $\gpara$, i.e., $\langle \Phi_{\vn k \gpara n} | \Phi_{\vn k^\prime \gpara^\prime m}\rangle =N_{\vn k}N_{\gpara} \delta_{\vn k \vn k^\prime}\delta_{\gpara \gpara^\prime}\delta_{nm}$, and they are periodic with respect to both $\vn k$ and $\gpara$.

The question arises to which Hamiltonian $\mathcal H$ the product states $\Phi_{\vn k \gpara n}(\vn r,\gpos)$, Eq.~\eqref{eq:product_states_end}, are eigenstates in the composite space $(\vn r, \gpos)$. Since the eigenstates have the product shape, the sought Hamiltonian decomposes into two additive contributions. If we denote by $\bar H$ the single-particle Hamiltonian to which the auxiliary orbital $\artpsi_{\gpara}$ is an eigenstate (see Eq.~\eqref{eq:art_SG}), the Hamiltonian of the composite system is given by
\begin{equation}
 \mathcal H(\vn r,\gpos) = H(\vn r) + \bar H(\gpos) \, .
\label{eq:gen_WF_H_comp}
\end{equation}
Here, the Hamiltonian $H$, which is independent of the parameter $\gpara$, can be written 
in the form
\begin{equation}
 H(\vn r) = \int H^{(\gpara)}(\vn r) \delta(\hat{\gpara}-\gpara) \,\D\gpara \,,
\end{equation}
where $\hat{\gpara}\Psi_{\vn k \gpara n} = \gpara \Psi_{\vn k \gpara n}$. When acting with this Hamiltonian on a specific Bloch state $\Psi_{\vn k \gpara n}$, the delta function selects the Hamiltonian $H^{(\gpara)}$ which corresponds to the specific parameter value of the Bloch state. It thus follows that $H \Psi_{\vn k \gpara n} = \mathcal{E}_{\vn k \gpara n} \Psi_{\vn k \gpara n}$ in line with Eq.~\eqref{eq:h_parameter}. Therefore, the product states satisfy the Schr\"odinger equation
\begin{equation}
 \mathcal H(\vn r,\gpos) \Phi_{\vn k \gpara n}(\vn r,\gpos) = \left( \mathcal{E}_{\vn k \gpara n} + \bar{\mathcal{E}}_{\gpara} \right)\Phi_{\vn k \gpara n}(\vn r,\gpos) \, ,
\label{eq:gen_WF_H_SG}
\end{equation}
where $\bar{\mathcal{E}}_{\gpara}$ represents the energy band in $\gpara$-space associated with the auxiliary orbital $\artpsi_{\gpara}$.

According to Eq.~\eqref{eq:gen_WF_H_SG}, the eigenvalues of the composite Hamiltonian $\mathcal H$ differ from the \abinitio{} band energies, which we would like to interpolate. To achieve the identity between the two sets of eigenvalues, we study the deep-well limit for the Hamiltonian $\bar H$. In this case, the energy level $\bar{\mathcal{E}}_{\gpara}$ becomes  independent of $\gpara$. As a consequence, $\bar{\mathcal{E}}_{\gpara}$ in Eq.~\eqref{eq:gen_WF_H_SG} can be set to zero, so that
\begin{equation}
 \mathcal H(\vn r,\gpos) \Phi_{\vn k \gpara n}(\vn r,\gpos) = \mathcal{E}_{\vn k \gpara n} \Phi_{\vn k \gpara n}(\vn r,\gpos) \, .
\end{equation}
Therefore, the generalized interpolation in $\vn k$ and $\gpara$ of the \bstruc{} of the composite Hamiltonian $\mathcal H$ grants access to the interpolated \bstruc{} of the physical Hamiltonian $H$ of interest.

\subsection{Higher dimensional Wannier functions (\gwfshort)}
\subsubsection{Definition}
Discrete Fourier transformations of the product states $\Phi_{\vn k \gpara m}$ with respect to $\vn k$ and $\gpara$ define \gwfshort{} in close analogy to MLWFs. We repeat here Eq.~\eqref{eq:gen_WF_def} as one of the main results of this work:
\begin{equation}
\begin{split}
 \gwann_{\vn R \gdlatt n}(\vn r,\gpos) = \frac{1}{N_{\vn k}}\frac{1}{N_{\gpara}} &\sum\limits_{\vn k \gpara m} \E{-\I \vn k\cdot \vn R} \E{-\I \gpara \cdot \gdlatt}\\
&\times \gumat_{mn}^{(\vn k,\gpara)} \Phi_{\vn k \gpara m}(\vn r, \gpos) \, .
\end{split}
\label{eq:gen_WF_def_close}
\end{equation}
\gwfshort{} are labeled by an orbital index $n$, the direct lattice vector $\vn R$, and an additional lattice vector $\gdlatt$. $\gdlatt$ is conjugate to $\gpara$ like the direct lattice vector $\vn R$ is conjugate to the crystal momentum $\vn k$. Unitary gauge transformations $\gumat^{(\vn k,\gpara)}$ control the localization of \gwfshort{}, and $N_{\vn k}$ and $N_{\gpara}$ stand for the number of grid points in $\vn k$-space and $\gpara$-space, respectively. Due to the orthogonality of the product states $\Phi_{\vn k \gpara m}$, the orbitals $\gwann_{\vn R \gdlatt n}(\vn r,\gpos)$ are orthonormal, i.e., $\langle W_{\vn R \gdlatt n} | W_{\vn R^\prime \gdlatt^\prime m}\rangle = \delta_{\vn R \vn R^\prime} \delta_{\gdlatt \gdlatt^\prime} \delta_{nm}$.

\subsubsection{Centers and spreads}
\label{subsec:centers}
A first physical interpretation of the functions $\gwann_{\vn R \gdlatt n}$ defined by Eq.~\eqref{eq:gen_WF_def_close} is provided by the expressions for the centers of \gwfshort{} in $\vn r$ and $\gpos$. The centers of \gwfshort{} in real space $\vn r$ can be directly related to the BZ sum of the Berry connection in crystal momentum space:
\begin{equation}
\label{eq:gen_WF_center_r}
\begin{split}
 \langle \gwann_{\vn 0 \vn 0 n} | \vn r | \gwann_{\vn 0 \vn 0 n} \rangle &= \frac{\I}{N_{\vn k}N_{\gpara}} \sum\limits_{\vn k \gpara} \langle \tilde \varphi_{\vn k \gpara n} | \nabla_{\vn k} | \tilde \varphi_{\vn k \gpara n} \rangle \\
 &= \frac{\I}{N_{\vn k}N_{\gpara}} \sum\limits_{\vn k \gpara} \langle \tilde u_{\vn k \gpara n} | \nabla_{\vn k} | \tilde u_{\vn k \gpara n} \rangle \, ,
 \end{split}
\end{equation}
which is easily derived using the Bloch-like periodic parts $\tilde \varphi_{\vn k \gpara n} = \sum_m\gumat_{mn}^{(\vn k,\gpara)} \varphi_{\vn k \gpara m}$ and $\tilde u_{\vn k \gpara n} = \sum_m\gumat_{mn}^{(\vn k,\gpara)} u_{\vn k \gpara m}$, respectively. Equation~\eqref{eq:gen_WF_center_r} is the generalization of the expression for centers of MLWFs. To obtain the $\gpos$-centers of \gwfshort{}, we start from the definition Eq.~\eqref{eq:gen_WF_def_close} and write down the expectation value of the auxiliary position operator in the basis of \gwfshort{}:
\begin{equation}
\begin{split}
 \langle \gwann_{\vn 0 \vn 0 n} | \gpos | \gwann_{\vn 0 \vn 0 n} \rangle &= \frac{\I}{N_{\vn k}N_{\gpara}} \sum\limits_{\vn k \gpara} \langle \tilde \varphi_{\vn k \gpara n} | \nabla_{\gpara} | \tilde \varphi_{\vn k \gpara n} \rangle \\
 &=\frac{\I}{N_{\vn k}N_{\gpara}} \sum\limits_{\vn k \gpara} \big( \langle \tilde u_{\vn k \gpara n} | \nabla_{\gpara} | \tilde u_{\vn k \gpara n} \rangle \\
 &\ \quad\qquad\qquad+ \langle \artu_{\gpara} | \nabla_{\gpara} | \artu_{\gpara} \rangle \big) \, .
\end{split}
\end{equation}
However, the second term $\langle \artu_{\gpara} | \nabla_{\gpara} | \artu_{\gpara} \rangle$ vanishes in the deep-well limit (see~e.g.~Eq.~\eqref{eq:auxiliary_solution}). Accordingly, the centers of \gwfshort{} in the auxiliary space $\gpos$ are given by the BZ sum of the Berry connections in $\gpara$-space:
\begin{equation}
 \langle \gwann_{\vn 0 \vn 0 n} | \gpos | \gwann_{\vn 0 \vn 0 n} \rangle = \frac{\I}{N_{\vn k}N_{\gpara}} \sum\limits_{\vn k \gpara} \langle \tilde u_{\vn k \gpara n} | \nabla_{\gpara} | \tilde u_{\vn k \gpara n} \rangle \, ,
\label{eq:gen_WF_center_xi}
\end{equation}
which are independent of the auxiliary orbitals $\artpsi_{\gpara}$ but determined solely by the Bloch-like periodic parts.

Likewise, the expectation values for the squared position operators $r^2$ and $\xi^2$ evaluate to
\begin{equation}
\begin{split}
 \langle \gwann_{\vn 0 \vn 0 n} | r^2 | \gwann_{\vn 0 \vn 0 n} \rangle &=\frac{-1}{N_{\vn k}N_{\gpara}} \sum\limits_{\vn k \gpara} \langle \tilde{\varphi}_{\vn k \gpara n} | \nabla_{\vn k}^2 | \tilde{\varphi}_{\vn k \gpara n} \rangle \\
 &= \frac{-1}{N_{\vn k}N_{\gpara}} \sum\limits_{\vn k \gpara} \langle \tilde u_{\vn k \gpara n} | \nabla_{\vn k}^2 | \tilde u_{\vn k \gpara n} \rangle \, ,
\end{split}
\label{eq:gen_WF_center_r2}
\end{equation}
and
\begin{equation}
\begin{split}
 \langle \gwann_{\vn 0 \vn 0 n} | \xi^2 | \gwann_{\vn 0 \vn 0 n} \rangle &=\frac{-1}{N_{\vn k}N_{\gpara}} \sum\limits_{\vn k \gpara} \langle \tilde{\varphi}_{\vn k \gpara n} | \nabla_{\gpara}^2 | \tilde{\varphi}_{\vn k \gpara n} \rangle \\
 &= \frac{-1}{N_{\vn k}N_{\gpara}} \sum\limits_{\vn k \gpara} \big( \langle \tilde u_{\vn k \gpara n} | \nabla_{\gpara}^2 | \tilde u_{\vn k \gpara n} \rangle \\ &\ \qquad\qquad\quad+\langle \artu_{\gpara}| \nabla_{\gpara}^2 | \artu_{\gpara}\rangle \big) \, .
\end{split}
\label{eq:gen_WF_center_xi2}
\end{equation}
While the $\gpos$-center is independent of the auxiliary orbital, the expectation value of $\xi^2$ contains explicitly a contribution from the integral $\langle \artu_{\gpara}| \nabla_{\gpara}^2 | \artu_{\gpara}\rangle$. Together with Eq.~\eqref{eq:gen_WF_center_r} and Eq.~\eqref{eq:gen_WF_center_xi} for the centers, the above expressions can be used to calculate the spread $\tilde \Omega$ of \gwfshort{} in the combined space of $\vn r$ and $\gpos$:
\begin{equation}
 \tilde\Omega = \sum\limits_n \left(\langle \gwann_{\vn 0 \vn 0 n}|\tilde r^2|\gwann_{\vn 0 \vn 0 n}\rangle - \langle \gwann_{\vn 0 \vn 0 n}|\vn{\tilde r}|\gwann_{\vn 0 \vn 0 n}\rangle^2\right)\, .
\label{eq:gen_WF_spread}
\end{equation}
To simplify notation, we introduced the generalized position operator $\vn{\tilde r}=(\vn r,\gpos)$.

\subsubsection{Maximal localization}
\label{subsec:construction_hdwfs}
As discussed in Sec.~\ref{sec:std_WFs}, the constraint of minimal spread $\Omega$ uniquely defines the MLWFs up to a global phase factor. Similarly, the unitary matrix $\gumat^{(\vn k, \gpara)}$ in the definition of \gwfshort{}, Eq.~\eqref{eq:gen_WF_def_close}, is determined from the condition that the orbitals $\gwann_{\vn R \gdlatt n}(\vn r, \gpos)$ should exhibit a minimal spread $\tilde \Omega$ in the space of $\vn r$ and $\gpos$. The resulting \gwfshort{} are unique up to a global phase factor.

Usually, the maximal localization procedure is performed in the three-dimensional real space $\vn r$. For constructing \gwfshort{}, we need to consider in addition the auxiliary space $\gpos$. To minimize the spread $\tilde\Omega$, Eq.~\eqref{eq:gen_WF_spread}, in the composite space of $\vn r$ and $\gpos$, we thus extend the \wprog{} program. Then, centers of \gwfshort{} possess additional coordinates, Eq.~\eqref{eq:gen_WF_center_xi}, owing to the auxiliary space. A higher dimensional but block diagonal Bravais matrix is employed to define the composite direct lattice in $(\vn r,\gpos)$-space:
\begin{equation}
 A = \begin{pmatrix} A_1 & 0 \\ 0 & A_2 \end{pmatrix} \, ,
\end{equation}
where $A_1$ is the usual $3\times 3$ Bravais matrix of the crystal and the rank of $A_2$ is given by the dimension of $\gpos$. Associated with the direct lattice is a composite reciprocal lattice in $(\vn k,\gpara)$-space combining crystal momentum and the additional parameter. Both $\vn k$ and $\gpara$ are chosen to form individual Monkhorst-Pack grids. According to the equations in Sec.~\ref{subsec:centers}, we can exploit the same finite-difference expressions as in Ref.~\cite{Marzari1997} to evaluate the spread $\tilde{\Omega}$. However, the role of the usual periodic parts $u_{\vn k n}(\vn r)$ is now taken by their higher dimensional analogs $\varphi_{\vn k \gpara n}(\vn r, \gpos)$, Eq.~\eqref{eq:product_periodic_end}. We need to set up all necessary neighbors $(\vn k+\vn b_{\vn k},\gpara+\vn b_{\gpara})$ of a point $(\vn k,\gpara)$ to apply the finite-difference formulas. Here, $\vn b_{\vn k}$ and $\vn b_{\gpara}$ connect the two reciprocal points. In general, we choose the Bravais matrix such that only those neighbors need to be considered where either $\vn b_{\vn k}=\vn 0$ or $\vn b_{\gpara}=\vn 0$.

The overlaps $\langle \varphi_{\vn k \gpara m} | \varphi_{\vn k+\vn b_{\vn k} \, \vn \gpara + \vn b_{\gpara} \, n}\rangle$ of the periodic parts at neighboring points in the $(\vn k, \gpara)$-space serve to calculate centers and spreads of \gwfshort{}. As we choose the directions of $\vn k$ and $\gpara$ to be orthogonal in the composite reciprocal lattice, the overlap matrix consists of the two contributions
\begin{align}
 M_{mn}^{(\vn k,\vn b)}(\gpara) &= \langle \varphi_{\vn k \gpara m} | \varphi_{\vn k+\vn b \, \gpara n}\rangle 
\label{eq:gen_WF_M_k_2}\\
 \bar M_{mn}^{(\gpara,\vn b)}(\vn k) &= \langle \varphi_{\vn k \gpara m} | \varphi_{\vn k \, \gpara+\vn b \, n}\rangle 
\label{eq:gen_WF_M_q_2}
\end{align}
depending on whether overlaps at neighboring $\vn k$-points or neighboring $\gpara$-points are concerned. The product shape of the periodic parts $\varphi_{\vn k \gpara m}$, Eq.~\eqref{eq:product_periodic_end}, allows further simplifications:
\begin{align}
 M_{mn}^{(\vn k,\vn b)}(\gpara) &= \langle u_{\vn k \gpara m} | u_{\vn k+\vn b \, \gpara n}\rangle \, ,
\label{eq:gen_WF_M_k}\\
\begin{split}
 \bar M_{mn}^{(\gpara,\vn b)}(\vn k) &= \langle u_{\vn k \gpara m} | u_{\vn k \, \gpara+\vn b \, n}\rangle \langle \artu_{\gpara} | \artu_{\gpara+\vn b}\rangle \\
 &= \mathcal M_{mn}^{(\gpara,\vn b)}(\vn k)\langle \artu_{\gpara} | \artu_{\gpara+\vn b}\rangle\, .
\label{eq:gen_WF_M_q}
\end{split}
\end{align}
The implementation of Eq.~\eqref{eq:gen_WF_M_k} within the FLAPW method is analogous to that of the usual overlaps, Eq.~\eqref{eq:overlap_k_std}, which is discussed in Ref.~\cite{Freimuth2008}. The evaluation of the overlaps in Eq.~\eqref{eq:gen_WF_M_q} requires the integrals $\langle \artu_{\gpara} | \artu_{\gpara+\vn b}\rangle$ in addition to the overlaps $\mathcal M_{mn}^{(\gpara,\vn b)}(\vn k)=\langle u_{\vn k \gpara m} | u_{\vn k \, \gpara+\vn b \, n}\rangle$ between the periodic parts of Bloch states. Details on the implementation of $\mathcal M_{mn}^{(\gpara,\vn b)}(\vn k)$ within the FLAPW method are given in the Appendices~\ref{app:details_FLAPW}-\ref{app:theta_FLAPW} for various realizations of $\gpara$. 

In addition, the projections of Bloch states onto localized trial orbitals $g_n(\vn r)$ are replaced by the projections
\begin{equation}
A_{mn}^{(\vn k,\gpara)} = \langle \Phi_{\vn k \gpara m} | p_n \rangle 
\label{eq:projections_gen}
\end{equation}
of the product states onto functions $p_n(\vn r, \gpos)$ localized in $(\vn r,\gpos)$-space. Such projections are the starting point for the spread minimization of \gwfshort{}. Exploiting the product shape of the states $\Phi_{\vn k \gpara m}$, Eq.~\eqref{eq:auxiliary_state}, we obtain
\begin{align}
 A_{mn}^{(\vn k,\gpara)} &= \langle \Psi_{\vn k \gpara m}| g_n \rangle\langle \artpsi_{\gpara}| h \rangle \, ,
\label{eq:gen_WF_A}
\end{align}
with the ansatz $p_n(\vn r,\gpos) = g_n(\vn r) h(\gpos)$. Thus, the projections onto localized trial functions factorize into the usual projections of Bloch states and the auxiliary projection $\langle \artpsi_{\gpara} | h \rangle$. In Appendix~\ref{app:details_FLAPW}, the construction of the usual projections $\langle \Psi_{\vn k \gpara m}| g_n \rangle$ within FLAPW is discussed.

\subsection{Generalized Wannier interpolation}
Within the energy window spanned by \gwfshort{}, the multi-parameter Hamiltonian $H^{(\vn k,\gpara)}$ can be interpolated in $\vn k$ and $\gpara$. As a starting point for the interpolation scheme, the matrix elements of the Hamiltonian $\mathcal H$ in the basis of \gwfshort{} have to be calculated:
\begin{equation}
\begin{split}
H_{nm}(\vn R,\gdlatt) &= \langle \gwann_{\vn 0 \vn 0 n} | \mathcal H | \gwann_{\vn R \gdlatt m} \rangle\\
&=\frac{1}{N_{\vn k}N_{\gpara}} \sum\limits_{\vn k \gpara n^\prime}\E{-\I \vn k \cdot \vn R}\E{-\I \gpara \cdot \gdlatt} \\
&\times \left(\gumat_{n^\prime n}^{(\vn k, \gpara)}\right)^* \mathcal{E}_{\vn k\gpara n^\prime}\,\gumat_{n^\prime m}^{(\vn k, \gpara)} \, .
\end{split}
\label{eq:gen_WF_hopp}
\end{equation}
These generalized hoppings are rapidly decaying with distance and, further, depend only on the distance vectors $\vn R^\prime - \vn R$ and $\gdlatt^\prime-\gdlatt$:
\begin{equation}
 \langle \gwann_{\vn R \gdlatt n} | \mathcal H | \gwann_{\vn R^\prime \gdlatt^\prime m} \rangle = \langle \gwann_{\vn 0 \vn 0 n} | \mathcal H | \gwann_{\vn R^\prime-\vn R \, \gdlatt^\prime-\gdlatt \, m} \rangle \, .
\end{equation}
The hoppings $H_{nm}(\vn R,\gdlatt)$ converge quickly with the number $N_{\vn k}\times N_{\gpara}$ of mesh points. They can therefore be constructed using a coarse $N_{\vn k}\times N_{\gpara}$ mesh. By an inverse Fourier transformation one obtains the interpolated $H^{(\vn k, \gpara)}$ for every desired point $(\vn k, \gpara)$, even if this point is not contained in the coarse $N_{\vn k}\times N_{\gpara}$ mesh used for the construction of the \gwfshort{}:
\begin{equation}
 H_{nm}^{(\vn k,\gpara)} =\sideset{}{'}\sum\limits_{\vn R\gdlatt}\E{\I \vn k\cdot \vn R}\E{\I \gpara\cdot \gdlatt} H_{nm}(\vn R,\gdlatt) \, .
\label{eq:ham_interpol}
\end{equation}
As the \gwfshort{} are strongly localized after minimizing their spread, the summation can be truncated. Indicated by the dashed symbol, only non-negligible hoppings are taken into account. Finally, the interpolated bands $\mathcal{E}_{\vn k\gpara n}$ are obtained by diagonalizing the interpolated Hamiltonian $H^{(\vn k, \gpara)}$:
\begin{equation}
 \left[\left(V^{(\vn k,\gpara)}\right)^\dagger H^{(\vn k,\gpara)} \, V^{(\vn k,\gpara)}\right]_{nm} = \mathcal{E}_{\vn k \gpara n}\delta_{nm} \, .
\label{eq:gen_WF_diag}
\end{equation}

\section{Application to spin spirals in a chain of \texorpdfstring{M\lowercase{n}}{Mn} atoms}
\label{sec:mn_chain}
\subsection{Heisenberg model and generalized Bloch theorem}
As an application of the generalized Wannier interpolation to realistic systems, we study a one-dimensional magnetic chain of Mn atoms oriented along the $z$ direction and extract Heisenberg exchange constants from \gwfshort{}. The Heisenberg model is defined as
\begin{equation}
 H = - \sum\limits_{ij} J_{ij} \vn S_i \cdot \vn S_j \, ,
 \label{eq:Heisenberg_H}
\end{equation}
where the Heisenberg exchange constants $J_{ij}$ mediate the exchange interaction between the normalized moments $\vn S_i$ and $\vn S_j$ located at the sites $i$ and $j$, respectively. In case of the magnetic monatomic chain, the most general solution to Eq.~\eqref{eq:Heisenberg_H} is the non-collinear (flat) spin-spiral state
\begin{equation}
 \vn S_n = (\cos naq, \sin naq, 0) \, ,
\label{eq:spin_spiral}
\end{equation}
which is characterized by the spin-spiral wave vector $\vn q = q \uvec_{z}$. Here, $a$ is the lattice constant along the chain axis. If we exploit translational invariance $J_{ij}=J_{0|j-i|}$ and Eq.~\eqref{eq:spin_spiral}, the energy of the system assumes the form
\begin{equation}
 E(q) = -2 \sum\limits_n J_{0n} \cos(naq) \, .
 \label{eq:E_tot}
\end{equation}
Expanding the energy in the vicinity of the ferromagnetic state ($q=0$), we can define the spin stiffness $A$ of the magnetic chain through $E(q\rightarrow 0) \approx E(0) + A q^2$. In order to access efficiently the Heisenberg exchange constants $J_{0n}$ as well as the spin stiffness $A$, we treat the spin-spiral vector $\vn q$ as an additional variable of a multi-parameter Hamiltonian $H^{(\vn k,\vn q)}$.

Without spin-orbit interaction we can make use of the so-called generalized Bloch theorem, which dictates a specific Bloch-like shape of the spinor eigenstates:
\begin{equation}
\Psi_{\vn k \vn q n}(\vn r) = \begin{pmatrix} \Psi^\uparrow_{\vn k \vn q n}(\vn r) \\ \Psi^\downarrow_{\vn k \vn q n}(\vn r) \end{pmatrix} = \E{i\vn k \cdot \vn r} \begin{pmatrix} \E{-i\frac{\vn q}{2}\cdot \vn r} \, u^\uparrow_{\vn k \vn q n}(\vn r) \\ \E{i\frac{\vn q}{2}\cdot \vn r} \, u^{\downarrow}_{\vn k \vn q n}(\vn r) \end{pmatrix}\, ,
\label{eq:gen_bloch_theorem}
\end{equation}
where $u^\uparrow_{\vn k \vn q n}(\vn r)$ and $u^\downarrow_{\vn k \vn q n}(\vn r)$ are lattice periodic functions. Using the latter ansatz allows us to avoid computationally demanding \firstprinciples{} calculations of large supercells and to perform all the calculations in a unit cell of one Mn atom. The Bloch states, Eq.~\eqref{eq:gen_bloch_theorem}, can be chosen to obey the periodic gauge in $\vn k$ and $\vn q$ simultaneously. However, we emphasize that due to the $\vn q$-dependent phases in Eq.~\eqref{eq:gen_bloch_theorem}, which arise from spin-$1/2$ rotation matrices, the period associated with the $\vn q$-mesh is enhanced by an overall factor of two (recall that spin-$1/2$ acquires a Berry phase of $\pi$ upon rotating by 360$^\circ$). Consequently, for the construction of \gwfshort{} we have to uniformly sample the range $[0,4\pi/a)$ of $\vn q$-values.

\subsection{Symmetry}
\label{subsec:symmetry}
We have mentioned above that the Bloch states, Eq.~\eqref{eq:gen_bloch_theorem}, are periodic on the interval $[0,4\pi/a)$ of $\vn q$-points. However, as we will show now, energy dispersion $\mathcal{E}_{\vn k \vn q n}$ and wave functions to a $\vn q$-value in $[2\pi/a,4\pi/a)$ can be derived from corresponding quantities in the interval $[0,2\pi/a)$. Therefore, the effective number of spin-spiral vectors at which the electronic structure needs to be calculated from \firstprinciples{} is reduced by a factor of two.

At a given spin-spiral vector $\vn q$, the Hamiltonian $H^{(\vn q)}$ has eigenvectors $\Psi_{\vn k \vn q n}$ and eigenvalues $\mathcal{E}_{\vn k \vn q n}$.
By symmetry, $H^{(\vn q)}$ is identical to the Hamiltonian $H^{(\vn q+\vn G)}$ at $\vn q+\vn G$, where $\vn G$ is a reciprocal lattice vector. Consequently, both Hamiltonians have (i) the same eigenvalue spectrum, i.e., $\mathcal{E}_{\vn k \vn q n} = \mathcal{E}_{\vn k^\prime \, \vn q+\vn G\,n}$  and (ii) the same set of eigenfunctions such that $\Psi_{\vn k \vn q n}=\Psi_{\vn k^\prime \, \vn q+\vn G\, n}$. In the \firstprinciples{} calculation, these eigenfunctions obey the generalized Bloch theorem, Eq.~\eqref{eq:gen_bloch_theorem}, which allows us to determine the above crystal momentum $\vn k^\prime$ explicitly:
\begin{equation}
\begin{split}
 \Psi_{\vn k \vn q n}(\vn r) &= \E{i\vn k \cdot \vn r} \begin{pmatrix} \E{-i\frac{\vn q}{2}\cdot \vn r} \, u^\uparrow_{\vn k \vn q n}(\vn r) \\ \E{i\frac{\vn q}{2}\cdot \vn r} \, u^{\downarrow}_{\vn k \vn q n}(\vn r) \end{pmatrix} \\ 
&=\E{i\vn k \cdot \vn r} \begin{pmatrix} \E{i\frac{\vn G}{2} \cdot \vn r}\,\E{-i\frac{\vn q+\vn G}{2}\cdot \vn r}  \, u^\uparrow_{\vn k \vn q n}(\vn r) \\ \E{i\frac{\vn G}{2} \cdot \vn r}\,\E{i\frac{\vn q+\vn G}{2}\cdot \vn r} \,\E{-i \vn G\cdot \vn r} u^{\downarrow}_{\vn k \vn q n}(\vn r) \end{pmatrix} \\
&=\E{i\left(\vn k +\frac{\vn G}{2}\right)\cdot \vn r} \begin{pmatrix} \E{-i\frac{\vn q+\vn G}{2}\cdot \vn r} \, \tilde u^\uparrow_{\vn k+\frac{\vn G}{2}\, \vn q+\vn G\, n}(\vn r) \\ \E{i\frac{\vn q+\vn G}{2}\cdot \vn r} \, \tilde u^{\downarrow}_{\vn k+\frac{\vn G}{2}\, \vn q+\vn G\, n}(\vn r) \end{pmatrix} \\
&= \Psi_{\vn k+\frac{\vn G}{2}\, \vn q+\vn G\, n}(\vn r) \, ,
\end{split}
\end{equation}
where we defined the lattice periodic functions
\begin{align}
\tilde u^\uparrow_{\vn k+\frac{\vn G}{2}\, \vn q+\vn G\, n}(\vn r)&=u^\uparrow_{\vn k \vn q n}(\vn r) \, ,
\label{eq:u_up_q}\\
\tilde u^{\downarrow}_{\vn k+\frac{\vn G}{2}\, \vn q+\vn G\, n}(\vn r)&=\E{-i\vn G \cdot \vn r} \, u^{\downarrow}_{\vn k \vn q n}(\vn r) \, .  
\label{eq:u_down_q}
\end{align}
If we change the spin-spiral vector $\vn q$ by $\vn G$, the Bloch state $\Psi_{\vn k \vn q n}$ and its energy $\mathcal{E}_{\vn k \vn q n}$ are moved to a different crystal momentum $\vn k^\prime = \vn k + \vn G/2$. Consequently, Bloch states need to be computed only for those spin-spiral vectors which lie in $[0,2\pi/a)$.

\subsection{Discussion of the implementation}
The evaluation of $M_{mn}^{(\vn k,\vn b)}(\gpara)$, Eq.~\eqref{eq:gen_WF_M_k}, does not differ from the case of standard MLWFs, except that $M_{mn}^{(\vn k,\vn b)}(\gpara)$ needs to be computed for several values of $\gpara=\vn q$. The matrix $\mathcal M_{mn}^{(\gpara,\vn b)}(\vn k)$ in Eq.~\eqref{eq:gen_WF_M_q} is given by the overlaps of periodic parts $u_{\vn k \vn q m}(\vn r)$ at neighboring spin-spiral vectors $\vn q$ and $\vn q+\vn b$. If we exploit the generalized Bloch theorem, Eq.~\eqref{eq:gen_bloch_theorem}, these overlaps assume the form
\begin{equation}
\begin{split}
 \mathcal M_{mn}^{(\vn q, \vn b)}(\vn k) &= \langle u_{\vn k \vn q m} | u_{\vn k\,\vn q+\vn b\,n}\rangle \\
&= \sum\limits_\sigma \int \E{\pm \I \frac{\vn b}{2}\cdot\vn r} \left(\Psi^\sigma_{\vn k \vn q m}(\vn r)\right)^* \Psi^\sigma_{\vn k\,\left[\vn q+\vn b\right]\,n}(\vn r)\,\D\vn r \, .
\end{split}
\label{eq:fleur_Mmn_maintext}
\end{equation}
Here, $[\vn q+\vn b]$ is a backfolding of the spin-spiral vector $\vn q+\vn b$ to the first BZ, and $\sigma=\uparrow,\downarrow$. The positive (negative) sign is taken in Eq.~\eqref{eq:fleur_Mmn_maintext} for the up component (down component) of the Bloch spinor. We describe in Appendix~\ref{app:details_FLAPW} the implementation of Eq.~\eqref{eq:fleur_Mmn_maintext} within the FLAPW method. To reduce the computational burden, we can apply the symmetry considerations of Sec.~\ref{subsec:symmetry} to the calculation of the above overlaps. We find that
\begin{equation}
 \mathcal M_{mn}^{(\vn q+\vn G, \vn b)}(\vn k) = \mathcal M_{mn}^{(\vn q, \vn b)}\left(\vn k+\frac{\vn G}{2}\right) = \mathcal M_{mn}^{(\vn q, \vn b)}\left(\vn k-\frac{\vn G}{2}\right) \, ,
\label{eq:symmetry_q}
\end{equation}
and likewise
\begin{equation}
 M_{mn}^{(\vn k, \vn b)}(\vn q+\vn G) = M_{mn}^{\left(\vn k+\frac{\vn G}{2}, \vn b\right)}(\vn q) = M_{mn}^{\left(\vn k-\frac{\vn G}{2}, \vn b\right)}(\vn q) \, ,
\label{eq:symmetry_k}
\end{equation}
where the periodic gauge of the Bloch states in $\vn k$-space was used. Thus, we can restrict ourselves to the calculation of $\mathcal M_{mn}^{(\vn q, \vn b)}(\vn k)$ and $M_{mn}^{(\vn k,\vn b)}(\vn q)$ for spin-spiral vectors in $[0,2\pi/a)$. Similarly, the projections $\langle \Psi_{\vn k \vn q m} | g_n \rangle$ in Eq.~\eqref{eq:gen_WF_A} need to be computed only for those $\vn q$ which lie in this interval.

Returning to Eq.~\eqref{eq:gen_WF_M_q}, we have to calculate additionally the auxiliary overlaps. Before, we modify the general shape of the auxiliary orbital $\artpsi_{\vn q}(\gpos)$, which was originally given by Eq.~\eqref{eq:auxiliary_state}. We choose $\artpsi_{\vn q}(\gpos) = \E{\I \frac{\vn q}{2} \cdot \gpos} \artu_{\vn q}(\gpos)$ such that the auxiliary orbital has the same $\vn q$-period as the Bloch states, Eq.~\eqref{eq:gen_bloch_theorem}. The lattice constant in the auxiliary space is thus given by $a$. Then, the auxiliary overlaps $\langle \artu_{\vn q}|\artu_{\vn q+\vn b}\rangle$ can be calculated numerically as discussed in Sec.~\ref{subsec:solution_orthogonality}. As an alternative, we can use the analytic expression Eq.~\eqref{eq:auxiliary_overlap} with $\tau = 2\pi/(N_{\vn q} a)$.

Projections $\langle \Psi_{\vn k \vn q m} | g_n \rangle$ and $\langle \artpsi_{\vn q}|h\rangle$ onto localized trial functions $g_n(\vn r)$ and $h(\gpos)$ enter Eq.~\eqref{eq:gen_WF_A}. The FLAPW implementation of the former is described in Appendix~\ref{app:details_FLAPW}. To obtain $\langle \artpsi_{\vn q}|h\rangle$, we project conveniently onto the single-well solution, Eq.~\eqref{eq:single_well_solution}, such that the integral is identical to one. However, projections onto different trial functions (e.g., Gaussians) can be employed as well. 

\subsection{Computational details}
Studies of $3d$ transition metal nanowires indicate that spin-orbit effects such as the magnetic anisotropy, which we discuss in Sec.~\ref{sec:mn_chain_mae}, should have a rather small influence on the electronic structure of Mn atoms due their half-filled $d$ shell \cite{Tung2007,Schubert2011}. Therefore, we neglect spin-orbit coupling for the moment. As first step, using the one-dimensional version \cite{Mokrousov2005} of the density functional theory J\"ulich FLAPW code \texttt{FLEUR} \cite{fleur}, we determine self-consistently the electronic density of a one-dimensional ferromagnetic linear chain of Mn atoms with a lattice constant of $a=5$\,bohr. We employ six local orbitals to treat the $3p$ core states of Mn. The RPBE parametrization of the exchange-correlation potential was used \cite{Zhang1998}. The non-overlapping muffin tin radii and the plane-wave cutoff were chosen to be $2.1$\,bohr and $3.8$\,bohr$^{-1}$, respectively. Starting from this charge density, we solve the Kohn-Sham equations on a uniform mesh of 8 $\vn k$-points separately for 16 spin-spiral vectors.

After that, the information about the wave functions at all $\vn k$- and $\vn q$-points is used to compute the necessary overlaps and projections. As first-guess trial orbitals $g_n$ we use three $d$ orbitals and six $sp^3d^2$ orbitals for each spin direction. The overlaps and projections of the auxiliary orbital $\artpsi_{\vn q}$ are derived either numerically or analytically as discussed before. A maximal real-space localization of the \gwfshort{} is achieved using our extension of the \wprog{} code to four space dimensions. Because of the metallic character of the magnetic Mn chain, a disentanglement \cite{Souza2001} of 18 optimally-connected quasi-Bloch states from a manifold of 36 Bloch orbitals is performed. The upper bound of the inner, or, frozen energy window is $2.2$ eV above the Fermi energy of the ferromagnetic state $E_F(q=0)$ (see Fig.~\ref{fig:Mn_disp}).

Constructing such \gwfshort{} requires a similar amount of computer time as the generation of individual sets of MLWFs for all of the 16 spin-spiral vectors. However, we emphasize that the single set of \gwfshort{} encodes the complete information of the electronic structure as a function of both $\vn k$ and $\vn q$ in the energy window of interest.

\subsection{Band structure interpolation}

\begin{figure}
 \centering
 \includegraphics{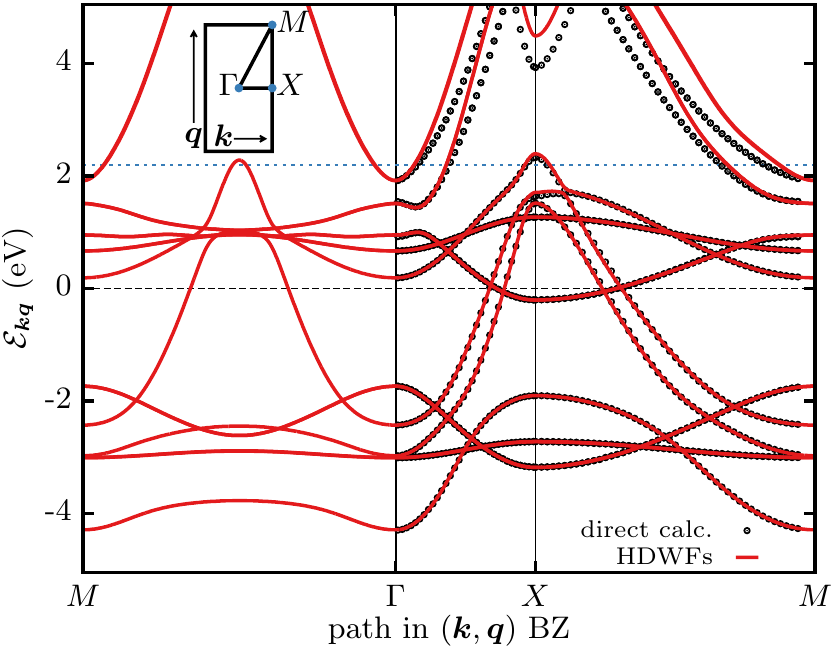}
 \caption{Generalized Wannier interpolation of the electronic \bstruc{} of a one-dimensional Mn chain along high-symmetry lines in the composite $(\vn k,\vn q)$ BZ. The interpolation based on \gwfshort{} (red solid lines) is in excellent agreement with direct first principles calculations (black circles). Energies are plotted relative to the Fermi level in the ferromagnetic case $E_F(q=0)$. The thin dotted line indicates the upper boundary of the inner energy window of $2.2$ eV.}
 \label{fig:Mn_disp}
\end{figure}

After constructing \gwfshort{} for the one-dimensional chain, we employ these functions in an interpolation of the multi-parameter Hamiltonian according to the discussion of Eq.~\eqref{eq:gen_WF_hopp}-Eq.~\eqref{eq:gen_WF_diag}. Figure~\ref{fig:Mn_disp} presents the results of the generalized Wannier interpolation of the band structure compared to the direct calculation. The inset of Fig.~\ref{fig:Mn_disp} does not show the usual BZ of crystal momentum but a composite BZ combining the crystal momentum $\vn k=k\uvec_z$ and the spin-spiral vector $\vn q=q\uvec_z$. According to the remark below Eq.~\eqref{eq:gen_bloch_theorem}, the composite BZ is given by $\left[-\pi/a,\pi/a\right]\times\left[-2\pi/a,2\pi/a\right]$, i.e., it is of rectangular shape.

A single set of 18 \gwfshort{} allows for an accurate interpolation of the energy bands in the reciprocal $(\vn k, \vn q)$-space. The path from the $\Gamma$-point ($k=0,q=0$) to the $X$-point ($k=\pi/a,q=0$) describes the electronic \bstruc{} of a ferromagnetic Mn chain. The standard Wannier interpolation for non-collinear or spin-spiral states, which was employed, e.g., in Ref.~\onlinecite{Hardrat2012}, is always restricted to high-symmetry lines parallel to $\Gamma-X$ where $q$ is constant. In contrast, the interpolation based on a single set of \gwfshort{} gives access to the electronic \bstruc{} along any given path in the composite BZ. Thereby we can easily compute the electronic \bstruc{} along the path from the $X$-point to the point $M$ ($k=\pi/a,q=2\pi/a$). Along $X-M$, the crystal momentum is kept fixed while the texture of the magnetic moments changes from the ferromagnetic ($q=0$) over to the antiferromagnetic state ($q=\pi/a$) and back to ferromagnetic order ($q=2\pi/a$). Due to the symmetry of the band structure discussed in Sec.~\ref{subsec:symmetry}, band energies differ at $X$ and $M$. The very same set of \gwfshort{} allows further for an interpolation of the \bstruc{} along the diagonal path $\Gamma-M$ of the BZ, which is not so easily accessible with the standard first principles codes. The band energies at $\Gamma$ and $M$ are identical due to symmetry (cf. Sec.~\ref{subsec:symmetry}). Overall, the accuracy of the generalized interpolation of the band structure is excellent within the frozen window.

\subsection{Real-space visualization of \gwfshort{}}

While MLWFs in magnetically collinear systems without spin-orbit coupling are real-valued \cite{Ri2014}, they are complex-valued functions in non-collinear systems, and in the presence of spin-orbit coupling \cite{Freimuth2008}. In contrast, the imaginary part of the \gwfshort{} of spin spirals is negligibly small such that we can restrict ourselves to a discussion of the real part of spinor valued \gwfshort{}.

In the following, we give an argument for the real-valuedness of \gwfshort{} for spin spirals in absence of spin-orbit coupling. We consider the Hamiltonian
\begin{equation}
\begin{split}
&H^{(\vn q)}(\vn r)= -\frac{\hbar^2}{2 m}\nabla^2 + V(\vn r) + \sum\limits_n B(\vn r- na\uvec_z) \vn S_n\cdot \vn\sigma \\
&= -\frac{\hbar^2}{2 m}\nabla^2 + V(\vn r) + \sum\limits_n B(\vn r- na\uvec_z)\begin{pmatrix}
                                                                              0 & \E{-\I n a q} \\
                                                                              \E{\I n a q} & 0
                                                                            \end{pmatrix} \, ,
\end{split}
\label{eq:H_ss}
\end{equation}
where the first two terms are the kinetic energy and the scalar potential. The last term describes the interaction with the noncollinear exchange field. The amplitude of the exchange field is given by $B(\vn r)$ and its direction is given by $\vn S_n$, Eq.~\eqref{eq:spin_spiral}, within the $n$-th atomic sphere.
It follows that
\begin{equation}
\left(H^{(\vn q)}(\vn r)\right)^* = H^{(-\vn q)}(\vn r) \, .
\label{eq:H_cc}
\end{equation}
If $\Psi_{\vn k \vn q n}$ is an eigenfunction of $H^{(\vn q)}$ to the real eigenvalue $\mathcal{E}_{\vn k \vn q n}$, we can apply a complex conjugation to the corresponding Schr\"odinger equation and arrive at
\begin{equation}
 H^{(-\vn q)}(\vn r) \left(\Psi_{\vn k \vn q n}(\vn r)\right)^* = \mathcal{E}_{\vn k \vn q n} \left(\Psi_{\vn k \vn q n}(\vn r)\right)^* \, ,
\end{equation} 
where Eq.~\eqref{eq:H_cc} was used. The complex conjugate of $\Psi_{\vn k \vn q n}$ is an eigenfunction of $H^{(-\vn q)}$ with energy $\mathcal{E}_{\vn k \vn q n}$. In general, eigenfunctions of $H^{(-\vn q)}$ are labeled by $\Psi_{\vn k -\vn q n}$ such that we necessarily need to find $(\Psi_{\vn k \vn q n})^* = \Psi_{\vn k^\prime  -\vn q n}$ for some crystal momenta $\vn k$ and $\vn k^\prime$. From the explicit shape of both states dictated by the generalized Bloch theorem, Eq.~\eqref{eq:gen_bloch_theorem}, follows that $\vn k^\prime = -\vn k$. We can choose the auxiliary orbital $\artpsi_{\vn q}$ of Eq.~\eqref{eq:auxiliary_state} to obey the relation $(\artpsi_{\vn q})^* = \artpsi_{-\vn q}$. Thus, the product states in Eq.~\eqref{eq:product_states_end} satisfy
\begin{equation}
\left(\Phi_{\vn k \vn q n}(\vn r,\gpos)\right)^* = \Phi_{-\vn k -\vn q n}(\vn r,\gpos) \, .
\label{eq:state_cc}
\end{equation}
If the unitary matrix satisfies
\begin{equation}
 \left(\gumat^{(\vn k,\vn q)}_{mn}\right)^* = \gumat^{(-\vn k,-\vn q)}_{mn} \, ,
\end{equation}
the real-valuedness of the \gwfshort{} is implied by Eq.~\eqref{eq:state_cc}:
\begin{equation}
\begin{split}
\gwann_{\vn 0 \vn 0 n} &= \frac{1}{N_{\vn k}N_{\gpara}}\sum\limits_{\vn k \vn qm}\gumat^{(\vn k,\vn q)}_{mn} \Phi_{\vn k \vn q m} \\
&= \frac{1}{2N_{\vn k}N_{\gpara}}\sum\limits_{\vn k \vn qm}\Big(\gumat^{(\vn k,\vn q)}_{mn}\Phi_{\vn k \vn q m} +\gumat^{(-\vn k,-\vn q)}_{mn}\Phi_{-\vn k -\vn q m}\Big) \\
&= \frac{1}{2N_{\vn k}N_{\gpara}}\sum\limits_{\vn k \vn qm}\Big(\gumat^{(\vn k,\vn q)}_{mn}\Phi_{\vn k \vn q m} + \left(\gumat^{(\vn k,\vn q)}_{mn}\Phi_{\vn k \vn q m}\right)^*\Big) \\
&= \frac{1}{N_{\vn k}N_{\gpara}}\Re\sum\limits_{\vn k \vn qm} \gumat^{(\vn k,\vn q)}_{mn}\Phi_{\vn k \vn q m} \, .
\end{split}
\label{eq:product_sum_real}
\end{equation}
A very similar argument shows that standard WFs can be chosen to be real-valued in some cases: For $\vn q=\vn 0$ Eq.~\eqref{eq:H_ss} describes a magnetically collinear system without spin-orbit coupling, for which Eq.~\eqref{eq:H_cc} and Eq.~\eqref{eq:state_cc} simplify into $\left(H(\vn r)\right)^*=H(\vn r)$ and $\left(\Psi_{\vn k n}(\vn r)\right)^*=\Psi_{-\vn k n}(\vn r)$, respectively. The choice $U_{mn}^{(-\vn k)} = (U_{mn}^{(\vn k)})^*$ in Eq.~\eqref{eq:WF_def} leads then to the real-valuedness of the resulting WFs.

\begin{figure}
 \centering
 \scalebox{0.18}{\includegraphics{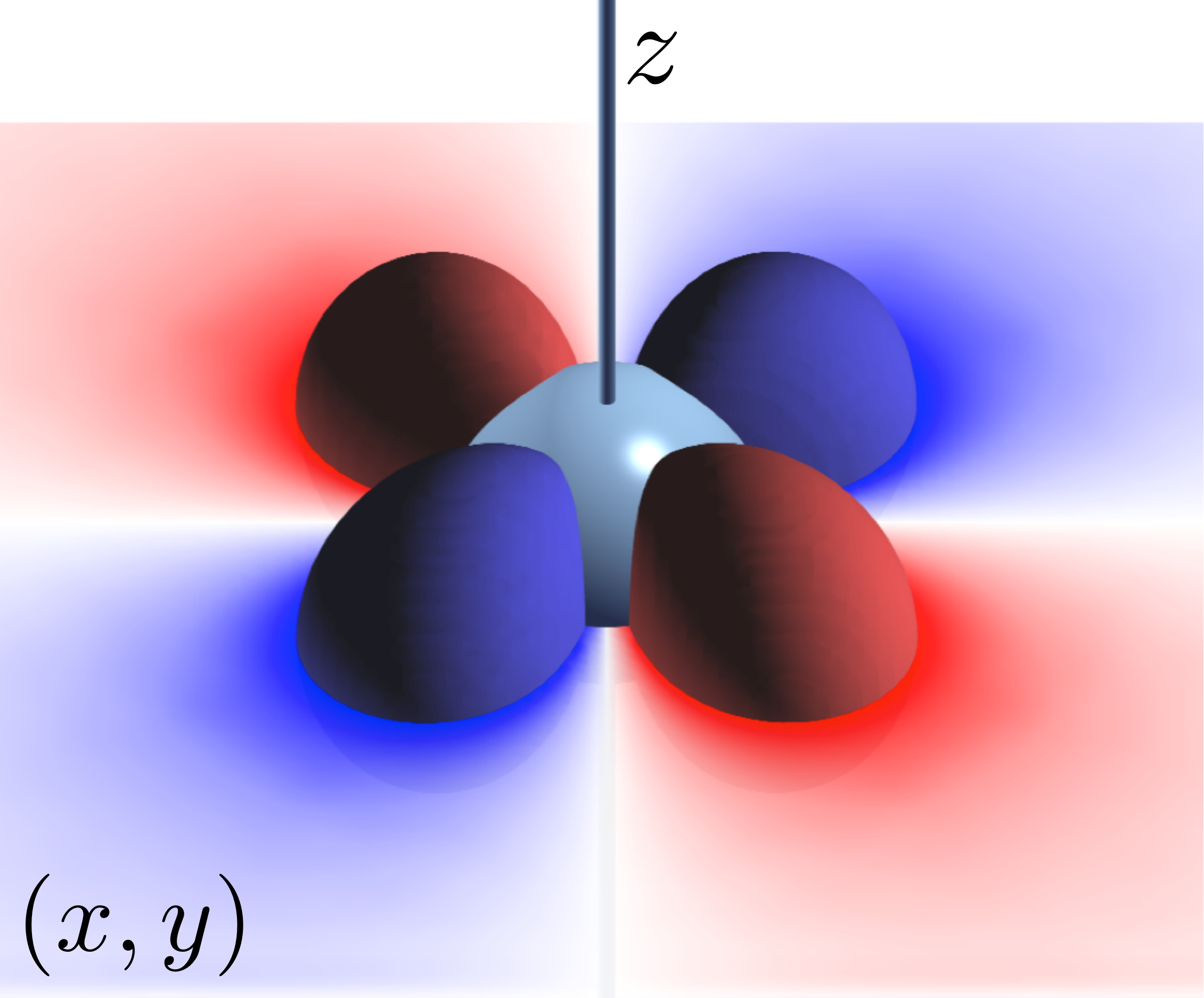}}\hfill
 \scalebox{0.18}{\includegraphics{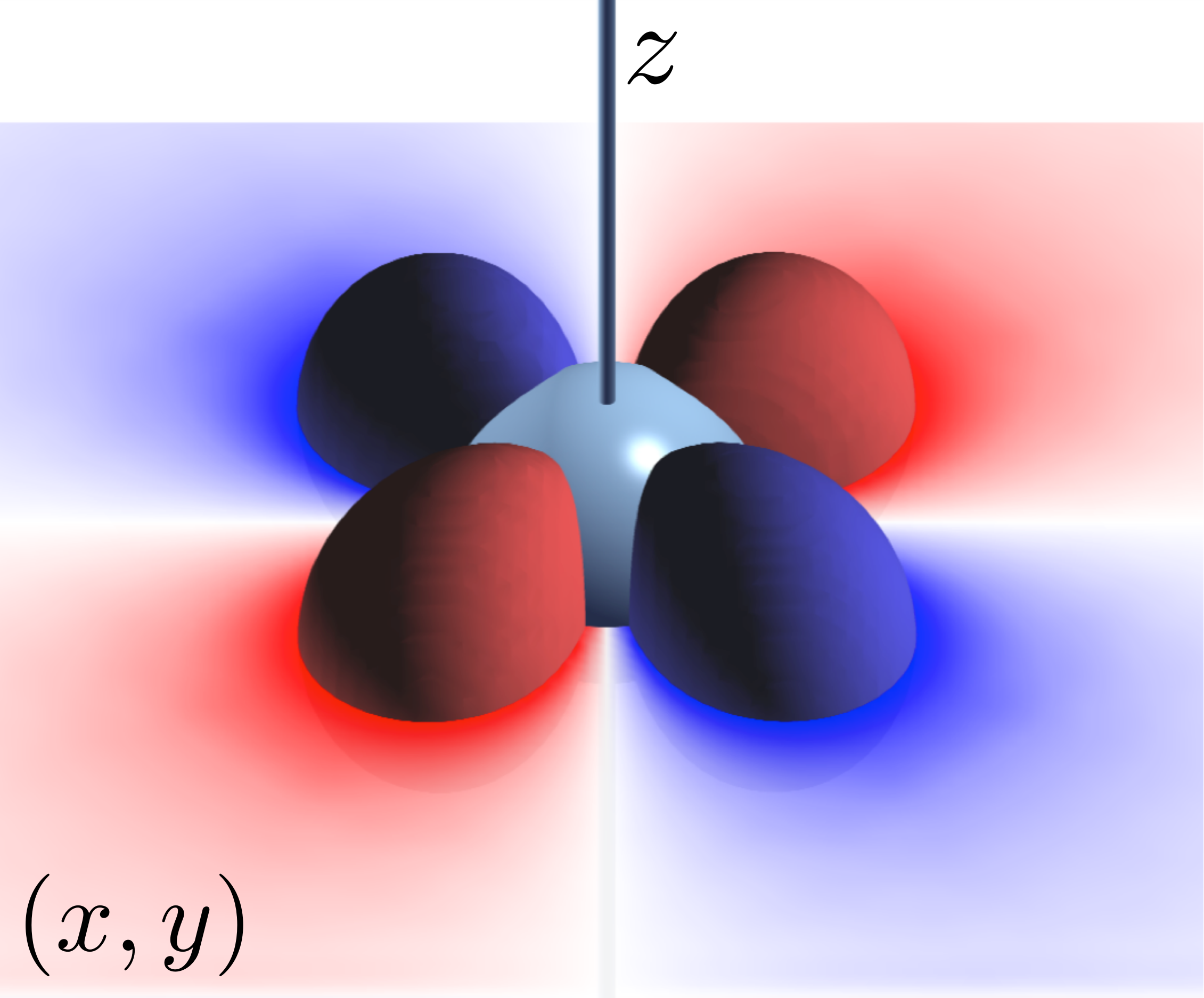}}
 \caption{Isosurfaces of a $d_{xy}$-like HDWF for $\xi_0=0$. We highlight the $xy$-plane which is perpendicular to the physical chain axis. The up-spin (left) is opposite in sign compared to the down-spin component (right), which is of equal magnitude. The functions were plotted using the program XCrySDen (Ref.~\onlinecite{Kokalj2003}).}
 \label{fig:fig_HDWFs_chain}
\end{figure}
\begin{figure}
 \centering
 \scalebox{0.18}{\includegraphics{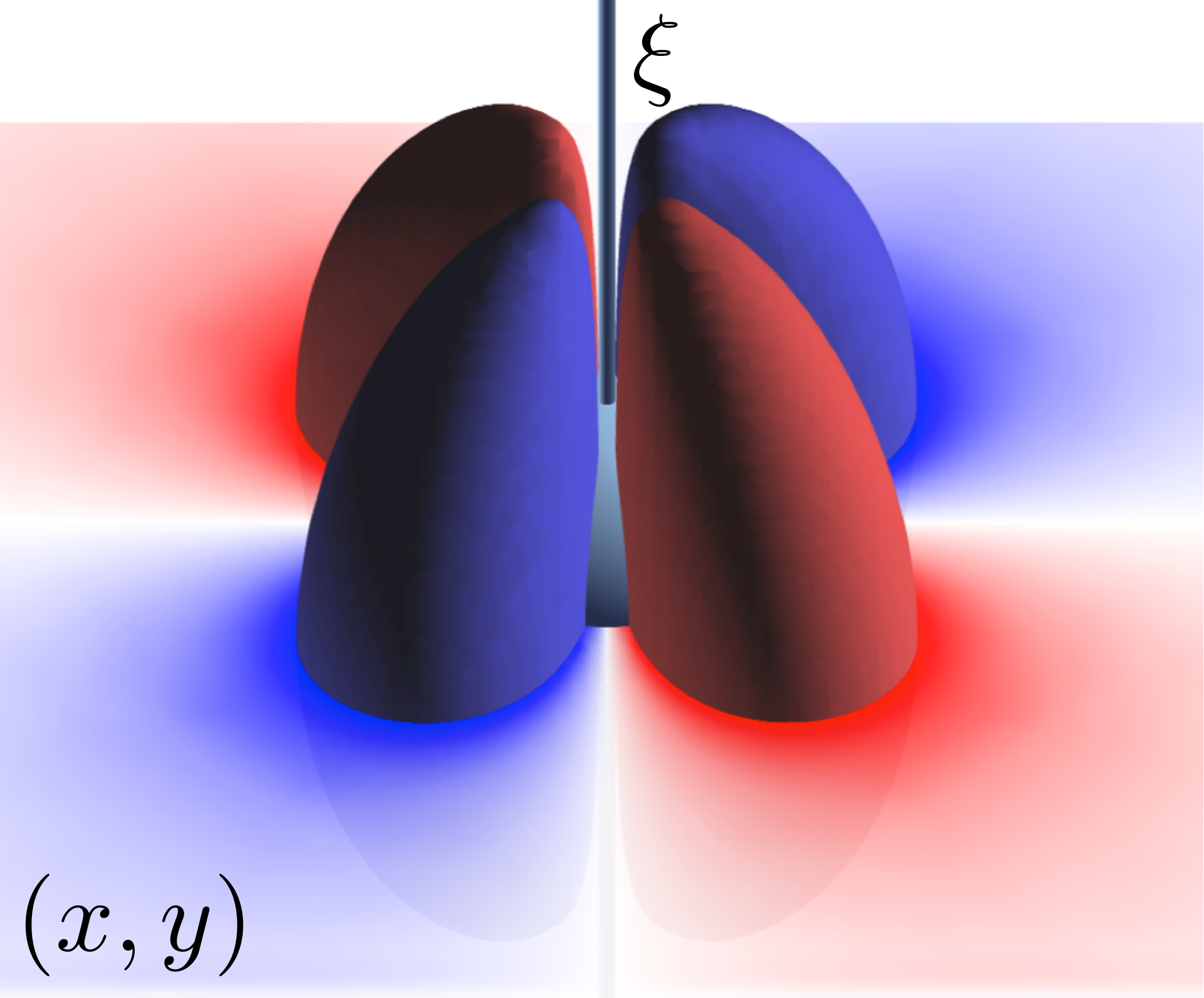}}\hfill
 \scalebox{0.18}{\includegraphics{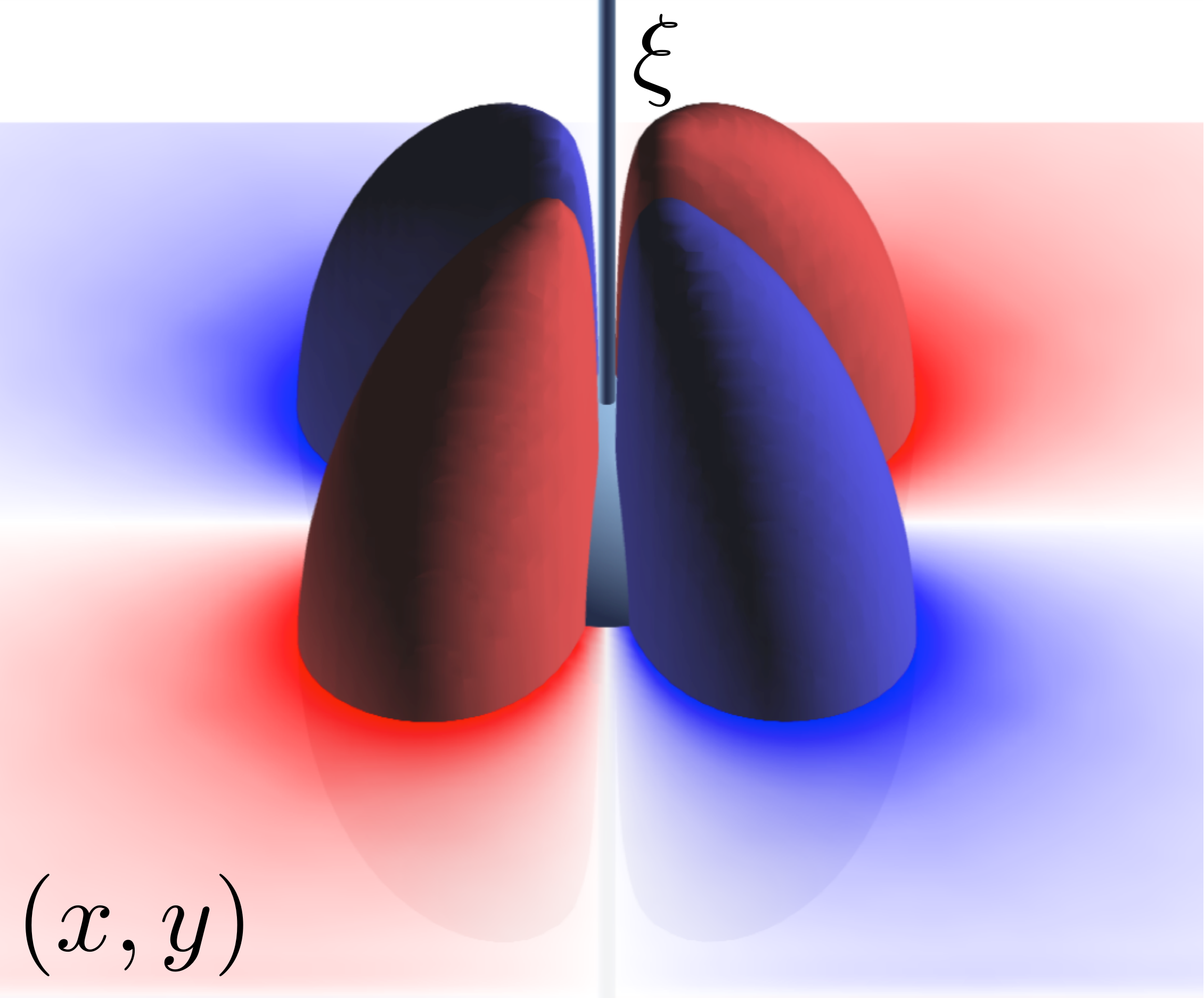}}
 \caption{Isosurfaces of a $d_{xy}$-like HDWF for $z_0=0$. The auxiliary dimension $\xi$ is perpendicular to the highlighted $xy$-plane. Up-spin component (left) and down-spin component (right) are of equal magnitude but opposite in sign.}
 \label{fig:fig_HDWFs_chain_2}
\end{figure}
To visualize \gwfshort{} in real space, we first divide a given HDWF by its phase at the maximal absolute value. Then, one of the four spatial coordinates ($x$, $y$, $z$ or $\xi$) is kept constant to obtain the three-dimensional plots of Fig.~\ref{fig:fig_HDWFs_chain} and Fig.~\ref{fig:fig_HDWFs_chain_2}. In general, we find that for a fixed auxiliary coordinate $\xi=\xi_0$, the \gwfshort{} (see Fig.~\ref{fig:fig_HDWFs_chain}) closely resemble the first-guess functions of $d$ and $sp^3d^2$ character visualizing the chemistry of the one-dimensional Mn chain. Variations of the value $\xi_0$ do not change the orbital character qualitatively. Choosing a constant value $z=z_0$ along the chain axis, we present the shape of a $d_{xy}$-like HDWF in Fig.~\ref{fig:fig_HDWFs_chain_2}. The HDWF extends throughout a single unit cell as a function of $\xi$ due to the construction of the auxiliary orbital, Eq.~\eqref{eq:auxiliary_state}, based on the deep-well limit.
\begin{figure}
 \centering
 \includegraphics{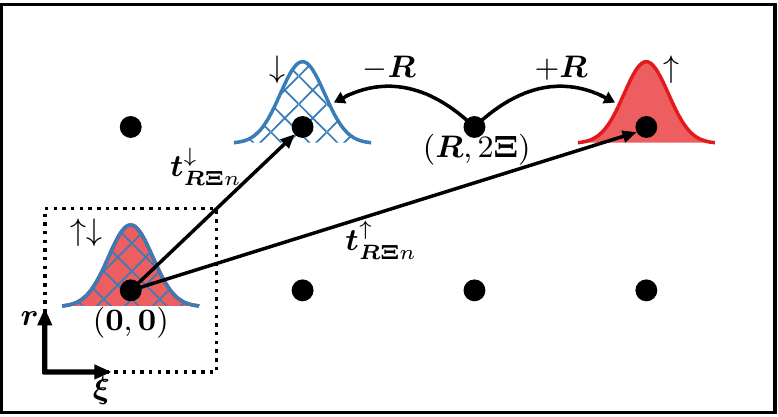}
 \caption{Scheme of the translation property Eq.~\eqref{eq:translations2} of \gwfshort{} in the composite lattice (black circles). Both spin components of the function $W_{\vn 0 \vn 0 n}$ are localized in the home unit cell $(\vn 0,\vn 0)$. In contrast, the spin components of $W_{\vn R \gdlatt n}$ are displaced in the auxiliary dimension with respect to the position $(\vn R,2\gdlatt)$. Arrows indicate the corresponding distance vectors $t^\uparrow_{\vn R \gdlatt n}$ and $t^\downarrow_{\vn R \gdlatt n}$.}
 \label{fig:fig_HDWFs_translation}
\end{figure}

In Fig.~\ref{fig:fig_HDWFs_chain} and Fig.~\ref{fig:fig_HDWFs_chain_2}, we show \gwfshort{} in the home unit cell, i.e., $\vn R=\vn 0$ and $\gdlatt=\vn 0$. How can we obtain the functions $W_{\vn R\gdlatt n}$ to finite $\vn R$ and $\gdlatt$ from those in the home unit cell? The spinor components are constructed according to
\begin{equation}
\begin{split}
 W^\sigma_{\vn R \gdlatt n}(\vn r,\gpos) = \frac{1}{N_{\vn k}N_{\vn q}}\sum\limits_{\vn k \vn q} &\E{-\I\vn k \cdot(\vn R-\vn r)} \E{\mp i\frac{\vn q}{2}\cdot \vn r} \E{-\I\vn q \cdot (\gdlatt-\frac{\gpos}{2})}\\
\times &\tilde u^\sigma_{\vn k \vn q n}(\vn r) \artu_{\vn q}(\gpos)
\end{split}
\label{eq:translations}
\end{equation}
which follows from Eq.~\eqref{eq:gen_WF_def_close}, Eq.~\eqref{eq:gen_bloch_theorem}, and the choice of the auxiliary orbital. Here, the Bloch-like periodic parts $\tilde u^\sigma_{\vn k\vn q n}=\sum_m \gumat^{(\vn k,\vn q)}_{mn} u^\sigma_{\vn k \vn q m}$ contain the unitary gauge matrix, and $\sigma=\uparrow,\downarrow$. In the case of the usual WFs, simple lattice translations need to be applied to obtain $W_{\vn R n}(\vn r)$ to any $\vn R$, i.e., $W_{\vn R n}(\vn r) = W_{\vn 0 n}(\vn r-\vn R)$. Compared to the usual WFs, we find from Eq.~\eqref{eq:translations} a slightly more complicated relation between $W_{\vn 0 \vn 0 n}(\vn r,\gpos)$ localized in the home unit cell and $W_{\vn R \gdlatt n}(\vn r,\gpos)$:
\begin{equation}
W^\sigma_{\vn 0 \vn 0 n}(\vn r,\gpos) = W^\sigma_{\vn 0 \gdlatt n}(\vn r,\gpos+2\gdlatt) = W^\sigma_{\vn R \vn 0 n}(\vn r+\vn R,\gpos\pm \vn R) \, ,
\end{equation}
and thus
\begin{equation}
W^\sigma_{\vn R \gdlatt n}(\vn r, \gpos) = W^\sigma_{\vn 0 \vn 0 n}(\vn r-\vn R,\gpos-2\gdlatt\mp\vn R)\, ,
\label{eq:translations2}
\end{equation}
where the upper (lower) sign is for the up (down) component of the spinor. We depict in Fig.~\ref{fig:fig_HDWFs_translation} the above translational property of HDWFs for spin spirals. Due to the coupling of $\vn k$ and $\vn q$ to the same real-space coordinate $\vn r$ (see Eq.~\eqref{eq:gen_bloch_theorem}), a spin-dependent shift of the spinor components occurs for finite $\vn R$. We can consider the distance vector between the centers of $W^\sigma_{\vn 0 \vn 0 n}(\vn r,\gpos)$ and $W^\sigma_{\vn R \gdlatt n}(\vn r,\gpos)$:
\begin{equation}
\begin{split}
 \vn t^\sigma_{\vn R \gdlatt n} &= \left\langle W^\sigma_{\vn R \gdlatt n} \left| \begin{pmatrix}\vn r \\ \gpos \end{pmatrix} \right| W^\sigma_{\vn R \gdlatt n} \right\rangle - \left\langle W^\sigma_{\vn 0 \vn 0 n} \left| \begin{pmatrix}\vn r \\ \gpos \end{pmatrix} \right| W^\sigma_{\vn 0 \vn 0 n} \right\rangle\\
&= \begin{pmatrix} \vn R \\ 2\gdlatt \pm \vn R \end{pmatrix} \, ,
\end{split}
\end{equation}
which follows from Eq.~\eqref{eq:translations2}. If $W^\sigma_{\vn 0 \vn 0 n}(\vn r,\gpos)$ is localized in the home unit cell at the position $(\vn r_c, \gpos_c)$, $W^\sigma_{\vn R \gdlatt n}(\vn r,\gpos)$ is centered at $(\vn r_c+\vn R, \gpos_c+2\gdlatt\pm\vn R)$ as shown in Fig.~\ref{fig:fig_HDWFs_translation}. While the direct lattice vector $\vn R$ determines the $\vn r$-center of the \gwfshort{} of spin spirals, the center in $\gpos$-space depends on both $\gdlatt$ and $\vn R$.

\subsection{Heisenberg exchange constants and spin stiffness}
\begin{figure}
 \centering
 \includegraphics{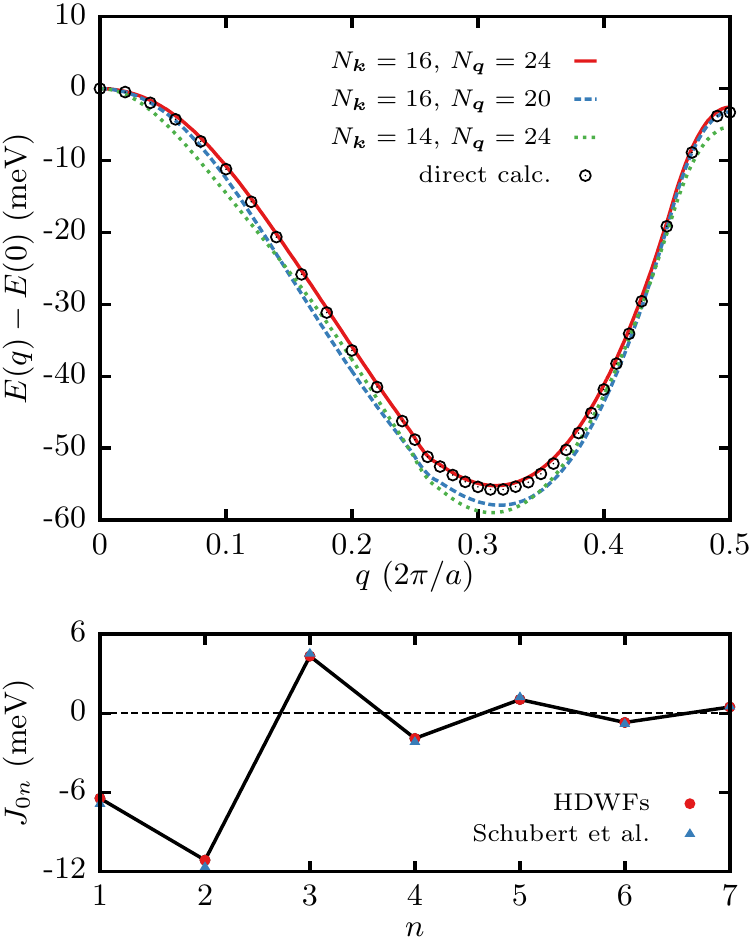}
 \caption{Top: Generalized interpolation of $E(q)-E(0)$ of the magnetic Mn chain. HDWF-interpolation (solid red line) reproduces the direct results (open circles) if the \gwfshort{} are constructed using $N_{\vn k}=16$ $\vn k$-points and $N_{\vn q}=24$ $\vn q$-points. Bottom: The Heisenberg exchange constants $J_{0n}$ obtained by fitting Eq.~\eqref{eq:E_tot} to the Wannier interpolated energy $E(q)$ (circles) are in excellent agreement with Ref.~\onlinecite{Schubert2011} (triangles).}
 \label{fig:fig_Jij}
\end{figure}

Starting from the generalized Wannier interpolation of the band structure throughout the $(\vn k, \vn q)$-space, we calculate the dispersion $E(q)$ of the system as the sum of occupied eigenvalues for a given value of $q$. Although the energy bands are interpolated nicely using a coarse mesh of 8 $\vn k$-points and 16 $\vn q$-points as shown in Fig.~\ref{fig:Mn_disp}, we find by comparison with direct first principles results (see upper panel of Fig.~\ref{fig:fig_Jij}) that an \abinitio{} mesh of 16 $\vn k$-points and 24 $\vn q$-points is necessary in order to interpolate the dispersion $E(q)$ of the system accurately. This means that Bloch functions need to be computed from first principles on a $(\vn k,\vn q)$-mesh of $16\times12$ if the symmetry relations from Sec.~\ref{subsec:symmetry} are exploited.

The value $q_0$ which minimizes the energy $E(q)$ defines the ground state of the magnetic system among all possible ferromagnetic, antiferromagnetic, and non-collinear spin-spiral configurations. Our interpolation of $E(q)$ in terms of \gwfshort{} reproduces precisely the curve obtained from direct calculation and thereby predicts the spin-spiral state with $q_0= 0.314\cdot 2\pi/a$ as the ground state of the one-dimensional Mn chain at the considered interatomic distance. The energy difference between ferromagnetic state and ground state amounts to $E(0)-E(q_0)= 55.2$ meV.
We further extract the Heisenberg exchange constants $J_{0n}$ by fitting Eq.~\eqref{eq:E_tot} to the HDWF-interpolated dispersion $E(q)$. The lower panel of Fig.~\ref{fig:fig_Jij} reveals that the exchange constants compare excellently with previous work \cite{Schubert2011}.

Wannier interpolation is particularly rewarding in the computation of transport properties as crystal momentum derivatives of the Hamiltonian can be taken analytically \cite{Usui2009,Shelley2011,Wang2006,Yates2007}. Such derivatives determine, for example, the velocity operator $\vn v=\nabla_{\vn k} H^{(\vn k)}/\hbar$. In the case of spin spirals, the derivative of $H^{(\vn k,\vn q)}$ with respect to $\vn q$ can be conveniently obtained from generalized Wannier interpolation:
\begin{equation}
 \frac{\partial H^{(\vn k,\vn q)}}{\partial q_\alpha} = \sum\limits_{\vn R \gdlatt} \I\Xi_\alpha \E{\I \vn k \cdot \vn R}\E{\I \vn q \cdot \gdlatt} H(\vn R,\gdlatt) \, .
\label{eq:deriv_H_q}
\end{equation}
Here, $H(\vn R,\gdlatt)$ is the matrix of the hoppings $H_{nm}(\vn R,\gdlatt)$, and $q_\alpha$ and $\Xi_\alpha$ refer to the $\alpha$-th components of the vectors $\vn q$ and $\gdlatt$, respectively. Such expressions allow us to calculate the second derivative of the energy $E(q)$ conveniently, from which we obtain the spin-stiffness $A$:
\begin{equation}
 A = \frac{1}{2}\frac{\partial^2 E(q)}{\partial q^2}\bigg|_{q=0} \, .
\label{eq:spin_stiff}
\end{equation}
Necessary details on the implementation of the scheme to obtain derivatives of $E(q)$ are provided in Appendix~\ref{app:derivative_H}. From the evaluation of Eq.~\eqref{eq:spin_stiff}, we obtain a value of $A = -174.1$ meV$\times$\AA$^2$ for the spin stiffness of the one-dimensional magnetic chain in the vicinity of the ferromagnetic state. To verify the estimated value for the spin stiffness, a polynomial even in $q$ is fitted to the interpolated dispersion near $q=0$. We extract a reference value of $-173.4$ meV$\times$\AA$^2$ from this fit, which is in very good agreement with the spin stiffness obtained from calculating directly Eq.~\eqref{eq:spin_stiff}.

\section{Application to the magnetic anisotropy in a chain of \texorpdfstring{M\lowercase{n}}{Mn} atoms}
\label{sec:mn_chain_mae}
\subsection{Introduction}
In this section, we discuss \gwfshort{} for the interpolation of the multi-parameter Hamiltonian $H^{(\vn k,\hat{\vn{m}})}$, where $\hat{\vn{m}}$ is the ferromagnetic magnetization direction. As an application, we consider the magneto-crystalline anisotropy energy (MAE), which is the energy difference between hard and easy axis of the system. Therefore, we adapt our description of the one-dimensional magnetic chain of Mn atoms to include spin-orbit coupling. The magnetization direction $\hat{\vn m}=(\sin\theta \cos \phi, \sin\theta \sin \phi, \cos \theta)$ is specified in spherical coordinates by $\theta$ and $\phi$. Here, we restrict ourselves to $\phi=0$. Bloch spinors and their periodic parts carry a dependence on $\theta$, i.e., $\Psi_{\vn k \theta n}(\vn r)=\E{\I\vn k \cdot \vn r}u_{\vn k \theta n}(\vn r)$. We include spin-orbit coupling by the second-variation scheme \cite{Li1990}.

\subsection{Discussion of the implementation}
According to Sec.~\ref{subsec:construction_hdwfs}, \gwfshort{} can be constructed using projections of the Bloch spinors and overlaps of their periodic parts. In the second-variation scheme, which we employ to include spin-orbit coupling, the coordinate system in spin space rotates together with $\hat{\vn m}$. Consequently, the spinors $u^{\phantom{\uparrow}}_{\vn k \theta n}(\vn r) = (u^\uparrow_{\vn k \theta n}(\vn r),u^\downarrow_{\vn k \theta n}(\vn r))$ and $u^{\phantom{\uparrow}}_{\vn k \,\theta+b\, n}(\vn r) = (u^\uparrow_{\vn k \,\theta+b\, n}(\vn r),u^\downarrow_{\vn k \,\theta+b\, n}(\vn r))$ refer to different spin-coordinate systems when $b\neq 0$. Thus, we need to transform the spinors into a common spin-coordinate frame when we compute the overlaps of the periodic parts at neighboring angles $\theta$ and $\theta+b$. We use the unitary rotation
\begin{equation}
 \chi(\theta) = \begin{pmatrix} \cos\frac{\theta}{2} & -\sin\frac{\theta}{2} \\[6pt] \sin\frac{\theta}{2} & \cos\frac{\theta}{2} \end{pmatrix}
 \label{eq:chi}
\end{equation}
to obtain all periodic parts $u_{\vn k \theta n}$ in the same global spin-coordinate frame by $u^{\text{gl}}_{\vn k \theta n}=\chi(\theta)u^{\phantom{\text{gl}}}_{\vn k \theta n}$. Then, the overlaps are given by
\begin{equation}
\begin{split}
\mathcal M_{mn}^{(\theta,b)}(\vn k) &= \langle u^{\text{gl}}_{\vn k \theta m} | u^{\text{gl}}_{\vn k\, \theta+b\, n}\rangle \\
&= \sum\limits_{\sigma\sigma^\prime} \left[\chi^\dagger(\theta) \chi(\theta+b)\right]_{\sigma\sigma^\prime} \langle u^{\sigma}_{\vn k \theta m} | u^{\sigma^\prime}_{\vn k\, \theta+b\, n}\rangle \, ,
\label{eq:Mmn_theta_maintext}
\end{split}
\end{equation}
where $\sigma=\uparrow,\downarrow$. The matrix elements $\langle u^{\sigma}_{\vn k \theta m} | u^{\sigma^\prime}_{\vn k\, \theta+b\, n}\rangle$ are
\begin{equation}
\langle u^{\sigma}_{\vn k \theta m} | u^{\sigma^\prime}_{\vn k\, \theta+b\, n}\rangle = \int \left(\Psi^\sigma_{\vn k \theta m}(\vn r)\right)^* \Psi^{\sigma^\prime}_{\vn k\,\left[\theta+b\right]\,n}(\vn r)\,\D\vn r \, ,
\label{eq:chain_temp_beta}
\end{equation}
where $[\theta+b]$ is a backfolding of the value $\theta+b$ to the one-dimensional BZ. These overlaps do not contain an additional $b$-dependent phase as in the cases of the spin spiral, Eq.~\eqref{eq:fleur_Mmn_maintext},  and standard MLWFs. We provide additional details and derive corresponding expressions to construct the matrix elements within the FLAPW method in Appendix~\ref{app:theta_FLAPW}. Apart from the overlaps $\mathcal M_{mn}^{(\theta,b)}(\vn k)$ we also need the overlaps $M_{mn}^{(\vn k,\vn b)}(\gpara)$ with $\gpara=\theta$ (see Eq.~\eqref{eq:gen_WF_M_k}). However, the calculation of $M_{mn}^{(\vn k,\vn b)}(\theta)$ does not differ from the case of standard MLWFs except that several values of $\theta$ need to be considered.

\subsection{Computational details}
We determine the charge density of the ferromagnetic chain self-consistently using the computational setup of Sec.~\ref{sec:mn_chain} but including now spin-orbit coupling in second-variation. Based on the converged electronic density, we solve on a uniform $\vn k$-mesh the Kohn-Sham equations for each magnetization direction separately. The values of $\theta$ are chosen from the range $[0,4\pi)$ as Bloch spinors acquire a minus sign upon $360^\circ$ rotation, i.e., $\Psi_{\vn k\, \theta+2\pi\, n}=-\Psi_{\vn k \theta n}$. However, symmetry considerations analogous to Sec.~\ref{subsec:symmetry} reduce the number of angles $\theta$ for which the Bloch functions need to be computed.

Then, the overlaps $\mathcal{M}^{(\theta,b)}_{mn}(\vn k)$, $M^{(\vn k,\vn b)}_{mn}(\theta)$ and projections are calculated. We project onto the same set of localized trial functions as in the case of Sec.~\ref{sec:mn_chain}, and further incorporate the analytical solution for the auxiliary orbital. After performing a disentanglement of 18 optimally-connected quasi-Bloch states from a manifold of 36 Bloch states with an inner window up to $2.2$ eV above the Fermi energy $E_F(\theta=0)$, we generate maximally localized \gwfshort{} using our extension of the \wprog{} program to four dimensions.

\subsection{Magnetic anisotropy}

\begin{figure}
 \centering
 \includegraphics{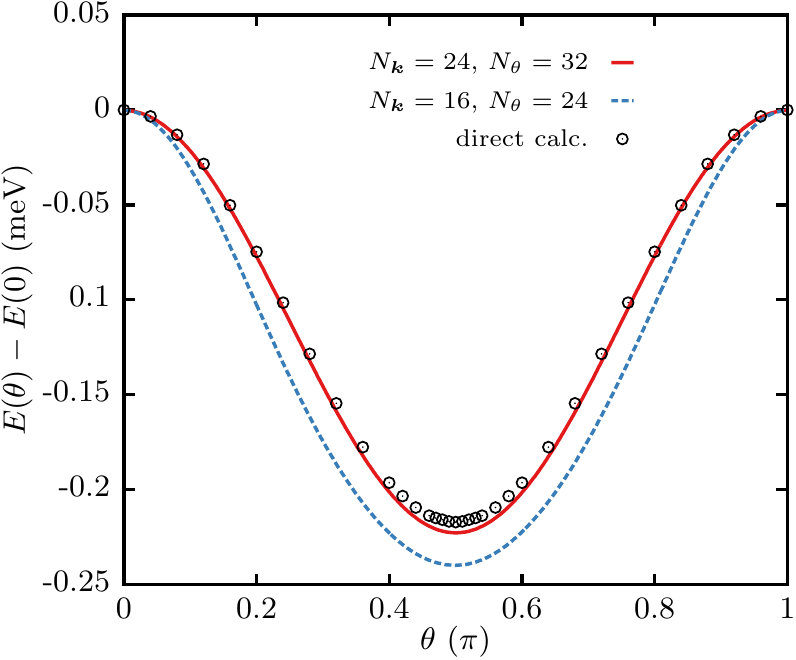}
 \caption{Generalized Wannier interpolation of $E(\theta)-E(0)$ of the magnetic Mn chain. Irrespective of the small energy scale, we can interpolate (solid red line) the energy difference as a function of the magnetization direction in very good agreement with the direct calculation (open circles).}
 \label{fig:fig_MAE}
\end{figure}

The single set of \gwfshort{} is employed to interpolate the energy bands $\mathcal{E}_{\vn k \theta n}$ in $\vn k$ and $\theta$ as described in Eq.~\eqref{eq:gen_WF_hopp}-Eq~\eqref{eq:gen_WF_diag}. In Fig.~\ref{fig:fig_MAE}, the energy difference $E(\theta)-E(0)$ is shown as obtained from such an energy interpolation. Compared to the spin-spiral application of Sec.~\ref{sec:mn_chain}, the \gwfshort{} have to be constructed on a denser \abinitio{} mesh of 24 $\vn k$- and 32 $\theta$-points (in $[0,4\pi)$) to reproduce the energy difference accurately. We associate this particularity with the small MAE of the one-dimensional Mn chain of $E(0)-E(\pi/2) = 0.217$ meV. Thus, a ferromagnetic magnetization direction perpendicular to the chain axis is favored over a parallel orientation as predicted in Ref.~\onlinecite{Tung2007}.

The uniaxial anisotropy energy can be parametrized by $E(\theta) = K_1 \sin^2 \theta$, where $K_1$ is the first anisotropy constant, and it follows that \cite{Wang1996}
\begin{equation}
 K_1 = \frac{\partial E(\theta)}{\partial \theta}\bigg|_{\theta=\pi/4} \, .
\label{eq:deriv_E_theta}
\end{equation}
Generalized Wannier interpolation can be employed conveniently to evaluate the above derivative with respect to the magnetization direction.
At zero temperature, we derive in Appendix~\ref{app:derivative_H} the expression
\begin{equation}
\begin{split}
 \frac{\partial E(\theta)}{\partial \theta}&=\frac{1}{N_{\vn k}}\sum\limits_{\mathcal{E}_{\vn k \theta n}\leq E_F(\theta)} \frac{\partial \mathcal{E}_{\vn k \theta n}}{\partial \theta} \\
 &=\frac{1}{N_{\vn k}}\sum\limits_{\mathcal{E}_{\vn k \theta n}\leq E_F(\theta)} \left\langle \varphi_{\vn k \theta n} \bigg| \frac{\partial H^{(\vn k,\theta)}}{\partial \theta} \bigg| \varphi_{\vn k \theta n} \right\rangle \, .
\end{split}
\label{eq:59}
\end{equation}
Here, $|\varphi_{\vn k \theta n}\rangle$ are eigenstates of $H^{(\vn k,\theta)}$ and the derivative of $H^{(\vn k,\theta)}$ can be obtained conveniently within the generalized Wannier interpolation scheme:
\begin{equation}
 \frac{\partial H^{(\vn k,\theta)}}{\partial \theta} = \sum\limits_{\vn R \Xi} \I\Xi \E{\I \vn k \cdot \vn R}\E{\I \theta\Xi} H(\vn R,\Xi) \, ,
\label{eq:deriv_H_theta}
\end{equation}
where $H(\vn R,\Xi)$ is the matrix of the hoppings $H_{nm}(\vn R,\Xi)$ between \gwfshort{}. The derivative of $E(\theta)$ at zero temperature, Eq.~\eqref{eq:59}, is the sum of the torques $\partial \mathcal{E}_{\vn k \theta n}/\partial \theta$ which electrons of band $n$ moving through the solid with crystal momentum $\vn k$ exert on the magnetization. Using this approach, we compute an anisotropy constant of $K_1 = -0.224$ meV, which agrees nicely with the value for $E(\pi/2)-E(0)$ given above.

\section{Possible further applications}
\label{sec:model}

Above, we have shown that \gwfshort{} can be constructed for \firstprinciples{} Hamiltonians and we discussed applications such as spin stiffness and MAE. In the following, we explore within models additional promising applications of \gwfshort{}.

\subsection{Virtual crystal approximation (VCA)}
The electronic structure of non-stoichiometric disordered alloys such as Fe$_{x}$Co$_{1-x}$ can be computed within VCA where virtual atoms with electronic structure corresponding to the concentration $x$ constitute a regular lattice \cite{Bellaiche2000}. The FLAPW method with a properly adjusted number of valence electrons is well-suited to describe these systems in case of alloys composed out of neighbors in the periodic table like Fe and Co \cite{Seemann2011}. By computing the electronic structure for several values of $x$, \gwfshort{} can be constructed which describe the multi-parameter Hamiltonian $H^{(\vn k, x)}$ for any concentration $x$. Thus, the treatment of alloys such as Fe$_{x}$Co$_{1-x}$ on a dense mesh of concentrations is simplified. The gauge of the alloy Hamiltonians is guaranteed to be smooth due to the single set of \gwfshort{} used in the generalized interpolation. If one simply mixes the MLWFs for $x=0$ and $x=1$, such a smooth gauge is more difficult to achieve \cite{Bianco2014}.

\begin{figure}
  \centering
  \includegraphics{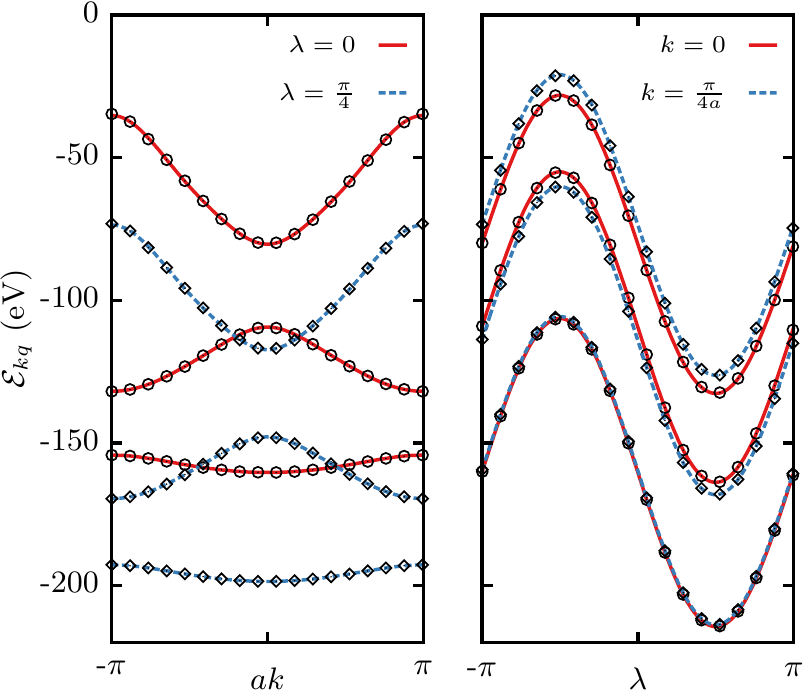}  
 \caption{Electronic \bstruc{} of the multi-parameter Hamiltonian Eq.~\eqref{eq:VCA_h} as obtained by generalized interpolation (solid and dashed lines). The dispersion is depicted as a function of the crystal momentum for constant $\lambda$ (left panel), and as a function of the parameter $\lambda$ for constant $k$ (right panel). The exact results (open circles and diamonds) agree nicely with the interpolation for the isolated group of energy bands. We have chosen the model parameters $a=3.0$~bohr, $b=2.9$~bohr, $\alpha=0.1$, and $V_0=544.0$~eV.}
 \label{fig:disp_vca}
\end{figure}

Here, we employ a toy model to outline the basic principle leaving the implementation into FLAPW for future work. We study modulations of the depth of attractive potential wells at positions $R_j = j a$ which define a one-dimensional lattice with lattice constant $a$ along the $z$ axis (see Fig.~\ref{fig:potential_well} for a sketch of the potential profile). The corresponding one-dimensional single-particle Hamiltonian carries a parametric dependence on the variable $\lambda$:
\begin{equation}
 H^{(\lambda)}(z) = -\frac{\hbar^2}{2m}\frac{\D^2}{\D z^2} -(1+\alpha\sin\lambda)V_0 \sum\limits_{R_j} \Theta_{R_j}^b(z) \, ,
\label{eq:VCA_h}
\end{equation}
where $|\alpha|<1$, and the well function $\Theta_{R_j}^b(z)$ is defined by Eq.~\eqref{eq:well_function}. The width $b$ and the potential strength $V_0$ are chosen such that the three lowest energy bands form an isolated group. The Hamiltonian of Eq.~\eqref{eq:VCA_h} is diagonalized in a plane-wave basis on a uniform $8\times 8$ $(k,\lambda)$-mesh. The $\lambda$-points lie in the interval $[0,2\pi)$. Necessary overlaps and projections onto localized Gaussians are constructed and the information on the auxiliary orbital $\zeta_{\lambda}$ is derived numerically as discussed in Sec.~\ref{subsec:solution_orthogonality}. Then, the \wprog{} program is used to achieve a maximal localization of \gwfshort{}. Figure~\ref{fig:disp_vca} demonstrates that the generalized interpolation reproduces the electronic \bstruc{} as a function of $k$ and $\lambda$ accurately.

\subsection{Ferroelectric polarization}
In ferroelectrics like the perovskite oxide BaTiO$_3$ \cite{Rabe2007}, a relative displacement characterized by the vector $\gpara$ of one of the sublattices leads to a change in ferroelectric polarization. To determine the
value of this change, the polarization in the form of MLWF centers or the Berry phase has to be computed along a certain path in $\gpara$-space \cite{King-Smith1993,Vanderbilt1993,Resta1994}. We can use the \gwfshort{} approach in order to interpolate the electronic structure of $H^{(\vn k,\gpara)}$ along the $\gpara$-path.

We use the following simple model to describe displacements between sublattices:
\begin{equation}
 H^{(\lambda)}(z) = -\frac{\hbar^2}{2m}\frac{\D^2}{\D z^2} - \sum\limits_{R_j}\bigg[V_0\Theta_{R_j}^b(z) 
+ V_0^\prime \Theta_{R_j}^b\left(z-\tau_\lambda\right)\bigg] \, ,
\label{eq:pol_h}
\end{equation}
where $\tau_\lambda=a/2+\delta_\lambda$ and $\delta_\lambda = -(b/2) \sin\lambda$ describes the relative displacement. In addition to the first well of depth $V_0$, which is kept fixed, the same unit cell contains a second well of strength $V_0^\prime$ at a variable position.
\begin{figure}
  \centering
  \includegraphics{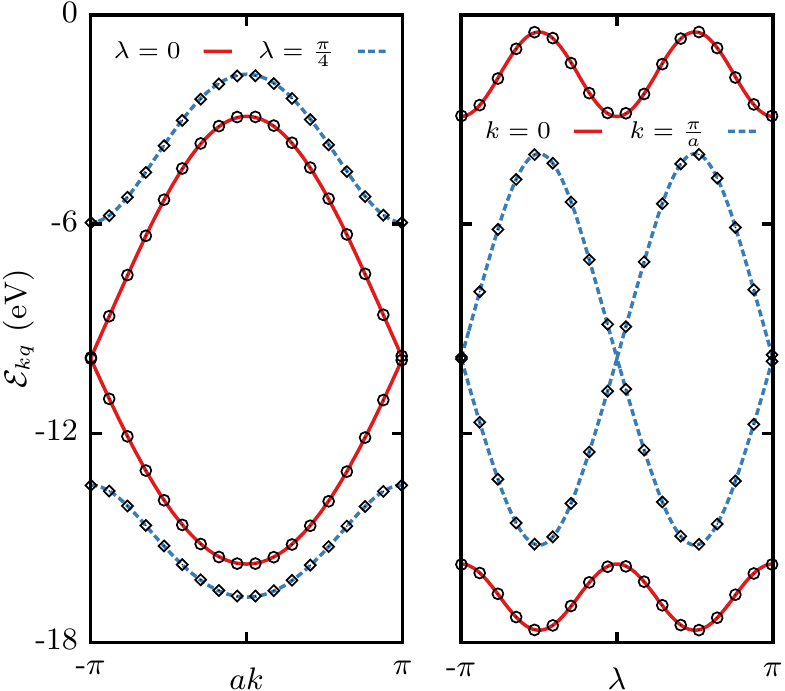}  
 \caption{Electronic \bstruc{} of the multi-parameter Hamiltonian Eq.~\eqref{eq:pol_h} as obtained by HDWF-interpolation (solid and dashed lines). We present the dispersion as a function of the crystal momentum at fixed $\lambda$ (left panel), and as a function of the parameter $\lambda$ at fixed $k$ (right panel). The interpolation is in excellent agreement with the exact results (open circles and diamonds). We have chosen $a=3.0$~bohr, $b=0.5$~bohr, and $V_0=V_0^\prime=272.0$~eV.}
 \label{fig:disp_distance}
\end{figure}
Again, we select the well parameters such that the two lowest bands form an isolated group. Employing a plane-wave basis, we diagonalize the Hamiltonian of Eq.~\eqref{eq:pol_h} on a mesh of $8$ $k$-points for each of the $8$ parameters $\lambda$ chosen uniformly from the range $[0,2\pi)$. Having at hand the Bloch states $\Psi_{k\lambda n}$, we construct overlaps and projections onto Gaussians. After deriving numerically the auxiliary orbital $\zeta_\lambda$ as discussed in Sec.~\ref{subsec:solution_orthogonality}, we use the \wprog{} code to establish a maximal localization of the \gwfshort{}. The \bstruc{} results of the generalized Wannier interpolation presented in Fig.~\ref{fig:disp_distance} are in excellent agreement with exact results. 

For a given value of $\gpara$ the ferroelectric polarization can be obtained as sum over the centers of WFs constructed from the occupied bands:
\begin{equation}
\vn P_{\gpara} = -\frac{e}{V}\sum\limits_{n\in\text{occ}} \langle W_{\vn 0 n}^{\gpara} | \vn r | W_{\vn 0 n}^{\gpara}\rangle \, .
\label{eq:pol_1}
\end{equation}
Here, $e>0$ is the positive electron charge, $V$ is the unit cell volume, and $|W_{\vn 0 n}^{\gpara}\rangle$ is a standard WF constructed according to Eq.~\eqref{eq:WF_def} for $H^{(\gpara)}$. The above ferroelectric polarization is not unique but defined up to the polarization quantum. Only polarization changes are unique and thus physical. However, to determine unambiguously the change of ferroelectric polarization between two points $\gpara_1$ and $\gpara_2$ according to Eq.~\eqref{eq:pol_1}, we usually need to ensure a smooth gauge of the Bloch states in $\gpara$-space. Such a gauge is guaranteed if we rewrite the ferroelectric polarization, Eq.~\eqref{eq:pol_1}, in terms of \gwfshort{}.

How can we now compute $\vn P_{\gpara}$ within the formalism of \gwfshort{}? By performing a Fourier transformation in $\gdlatt$, we obtain WFs at $\gpara$:
\begin{equation}
\begin{split}
\sum\limits_{\gdlatt} \E{\I \gpara\cdot \gdlatt} \gwann_{\vn R \gdlatt n}(\vn r,\gpos) &= \frac{1}{N_{\vn k}} \sum\limits_{\vn k m} \E{-\I \vn k \cdot \vn R} \gumat_{mn}^{(\vn k,\gpara)} \Phi_{\vn k \gpara m}(\vn r,\gpos) \\
&= \frac{\artpsi_{\gpara}(\gpos)}{N_{\vn k}}\sum\limits_{\vn k m} \E{-\I \vn k \cdot \vn R} \gumat_{mn}^{(\vn k,\gpara)} \Psi_{\vn k \gpara m}(\vn r) \\
&= \artpsi_{\gpara}(\gpos)W_{\vn R n}^{\gpara}(\vn r) \, .
\end{split}
\label{eq:pol_2}
\end{equation}
Recall that we generate just a single set of \gwfshort{} encoding the electronic structure information in $(\vn k,\gpara)$-space. Consequently, a smooth gauge is automatically built into the construction of \gwfshort{} such that also the above standard WFs are guaranteed to be smooth in $\gpara$. Inserting Eq.~\eqref{eq:pol_2} into Eq.~\eqref{eq:pol_1} yields
\begin{equation}
\begin{split}
 \vn P_{\gpara} &= -\frac{e}{VN_{\gpara}}\sum\limits_{n\in\text{occ}}\sum\limits_{\gdlatt{\gdlatt}^\prime} \E{\I \gpara\cdot({\gdlatt}^\prime-\gdlatt)} \langle \gwann_{\vn 0 \gdlatt n} | \vn r | \gwann_{\vn 0 {\gdlatt}^\prime n}\rangle \\
&=-\frac{e}{V}\sum\limits_{n\in\text{occ}}\sum\limits_{\gdlatt} \E{\I \gpara\cdot\gdlatt} \langle \gwann_{\vn 0 \vn 0 n} | \vn r | \gwann_{\vn 0 \gdlatt n}\rangle \, ,
\end{split}
 \label{eq:pol_3}
\end{equation}
where we exploited
\begin{equation}
\langle W_{\vn 0 n}^{\gpara}| \vn r| W_{\vn 0 n}^{\gpara}\rangle = \frac{1}{N_{\gpara}}\langle \artpsi_{\gpara}W_{\vn 0 n}^{\gpara}| \vn r | \artpsi_{\gpara}W_{\vn 0 n}^{\gpara}\rangle \, ,
\end{equation}
which follows from $\langle \artpsi_{\gpara}|\artpsi_{\gpara}\rangle=N_{\gpara}$. Equation~\eqref{eq:pol_3} can be employed to obtain an interpolated value for the ferroelectric polarization at values $\gpara$ that lie between the points of the coarse $\gpara$-mesh used to generate the \gwfshort{}. Of course, this works only if the system is insulating along the entire considered $\gpara$-path. The $\gpara$-sum of $\vn P_{\gpara}$ evaluates to
\begin{equation}
\begin{split}
\sum\limits_{\gpara} \vn P_{\gpara} &=
-\frac{e}{V}\sum\limits_{n\in\text{occ}}\sum\limits_{\gdlatt}\sum\limits_{\gpara}\E{\I \gpara\cdot\gdlatt} \langle \gwann_{\vn 0 \vn 0 n} | \vn r | \gwann_{\vn 0 \gdlatt n}\rangle \\
&=-\frac{eN_{\gpara}}{V}\sum\limits_{n\in\text{occ}}\langle \gwann_{\vn 0 \vn 0 n} | \vn r| \gwann_{\vn 0 \vn 0 n}\rangle \, ,
\end{split}
\end{equation}
which is determined by the centers of HDWFs available in the extended \wprog{} implementation.

Analogously to derivatives of the multi-parameter Hamiltonian discussed in Appendix~\ref{app:derivative_H}, we can calculate $\gpara$-derivatives of the ferroelectric polarization, Eq.~\eqref{eq:pol_3}, which read
\begin{equation}
 \frac{\partial \vn P_{\gpara}}{\partial \lambda_{\alpha}} = -\frac{e}{V}\sum\limits_{n\in\text{occ}}\sum\limits_{\gdlatt} \I \Xi_{\alpha} \E{\I \gpara\cdot\gdlatt} \langle \gwann_{\vn 0 \vn 0 n} | \vn r | \gwann_{\vn 0 \gdlatt n}\rangle\, .
\end{equation}
Here, $\lambda_{\alpha}$ and $\Xi_{\alpha}$ are the $\alpha$-th components of $\gpara$ and $\gdlatt$, respectively. Differentiating the ferroelectric polarization with respect to the sublattice displacement $\vn \delta$ (which depends on $\gpara$), we obtain the Born effective charge tensor. For the one-dimensional model of Eq.~\eqref{eq:pol_h} it follows that
\begin{equation}
\begin{split}
 Q^B&=\left.\frac{\partial P_{\delta}}{\partial \delta}\right|_{\delta= 0} = \left.\frac{\partial P_{\lambda}}{\partial \lambda}\frac{\partial\lambda}{\partial \delta}\right|_{\lambda=0} = -\frac{2}{b} \left.\frac{\partial P_{\lambda}}{\partial \lambda}\right|_{\lambda=0} \\
&= \frac{2e}{bV}\sum\limits_{n\in\text{occ}}\sum\limits_{\Xi} \I \Xi \langle \gwann_{0 0 n} | z | \gwann_{0 \Xi n}\rangle \, .
\end{split}
\end{equation}

\subsection{Current-induced torques in noncollinear magnetic systems}
\label{subsec:torques}
Current-induced torques on the magnetization (spin torques) are thought to play an important role in future magnetic memory devices. These spin torques result from the exchange of angular momentum between two magnets of distinct magnetization direction (spin transfer torques) \cite{Berger1996,Slonczewski1996,Sun1999}, or between spin and lattice (spin-orbit torques) \cite{Freimuth2014a,Garello2013,Chernyshov2009,Miron2010,Miron2011}. The spin-orbit torques can depend strongly on the magnetization direction \cite{Garello2013}. We expect that \gwfshort{} provide a convenient scheme to extract this directional dependence. 

To demonstrate that the generalized interpolation of the Hamiltonian with respect to isolated spin moment rotations in real space can be tackled with the formalism of HDWFs, we modify Eq.~\eqref{eq:VCA_h} to describe two magnetic ``atoms"  separated by half the lattice constant $s=a/2$. And while we keep the orientation of one of the atoms fixed, i.e., $\hat{\vn n}_1 = \uvec_{x}$, the direction of the other moment, $\hat{\vn n}_2 = \left(\cos \lambda,\sin\lambda,0\right)$, is tilted by the angle $\lambda$. 
\begin{figure}
  \centering
  \includegraphics{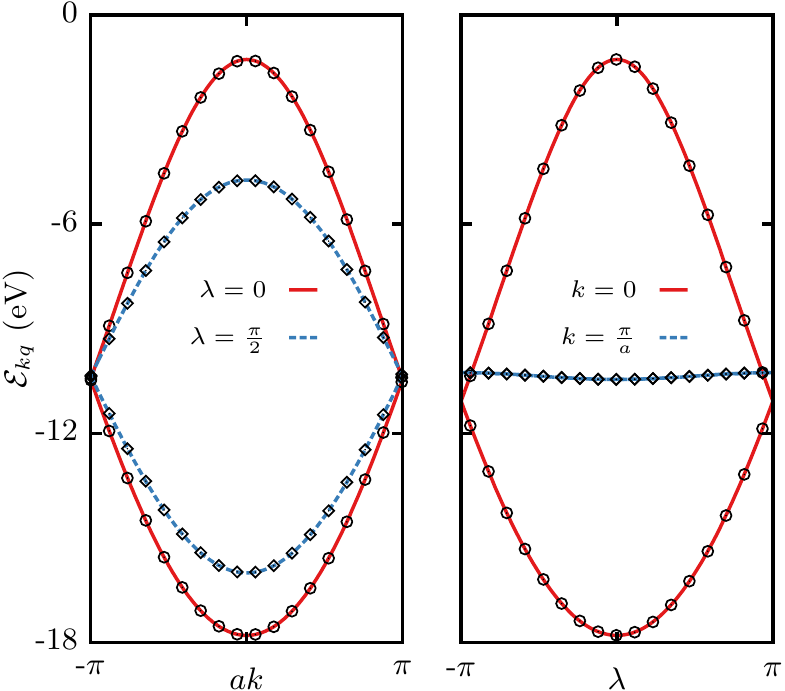}  
 \caption{Treating the tilting angle $\lambda$ as additional variable in the multi-parameter Hamiltonian, we can interpolate (solid and dashed lines) accurately the electronic \bstruc{} throughout $(k,\lambda)$-space. The dispersion is depicted either as a function of the crystal momentum for constant $\lambda$ (left panel), or as a function of the parameter $\lambda$ for constant $k$ (right panel). Open circles and diamonds indicate the exact results. The model parameters are $a=3.0$~bohr, $b=1.0$~bohr, $V_0=272.0$~eV, and $B_0=27.2$~eV.}
 \label{fig:disp_stt}
\end{figure}
The resulting single-particle Hamiltonian assumes the form:
\begin{equation}
\begin{split}
 H^{(\lambda)}(z) = &-\frac{\hbar^2}{2m}\frac{\D^2}{\D z^2} - V_0\sum\limits_{R_j}\left[\Theta_{R_j}^b(z) + \Theta_{R_j}^b(z-s)\right] \\
 &+B_0\sum\limits_{R_j}\left[\Theta_{R_j}^b(z) \hat{\vn n}_1 + \Theta_{R_j}^b(z-s) \hat{\vn n}_2\right]\cdot \vn\sigma  ,
\end{split}
\label{eq:stt_h}
\end{equation}
where $B_0$ is the strength of the exchange potential, and $\vn\sigma$ is the vector of Pauli matrices. Diagonalizing the matrix Eq.~\eqref{eq:stt_h} in a plane-wave basis on a coarse $8\times 16$ $(k,\lambda)$-mesh allows us to extract the Bloch spinors, and to calculate the matrices necessary to apply the \wprog{} minimization to \gwfshort{}. Recalling that the Bloch spinors acquire a Berry phase of $\pi$ upon rotating by $360^\circ$, we have to choose the $\lambda$-points uniformly in the interval $[0,4\pi)$ (see also the remark below Eq.~\eqref{eq:gen_bloch_theorem}). The auxiliary orbital is taken as $\artpsi_\lambda(\xi)=\E{\I\frac{\lambda}{2}\xi}\artu_\lambda(\xi)$.

\begin{figure}
  \centering
     \scalebox{0.35}{\includegraphics{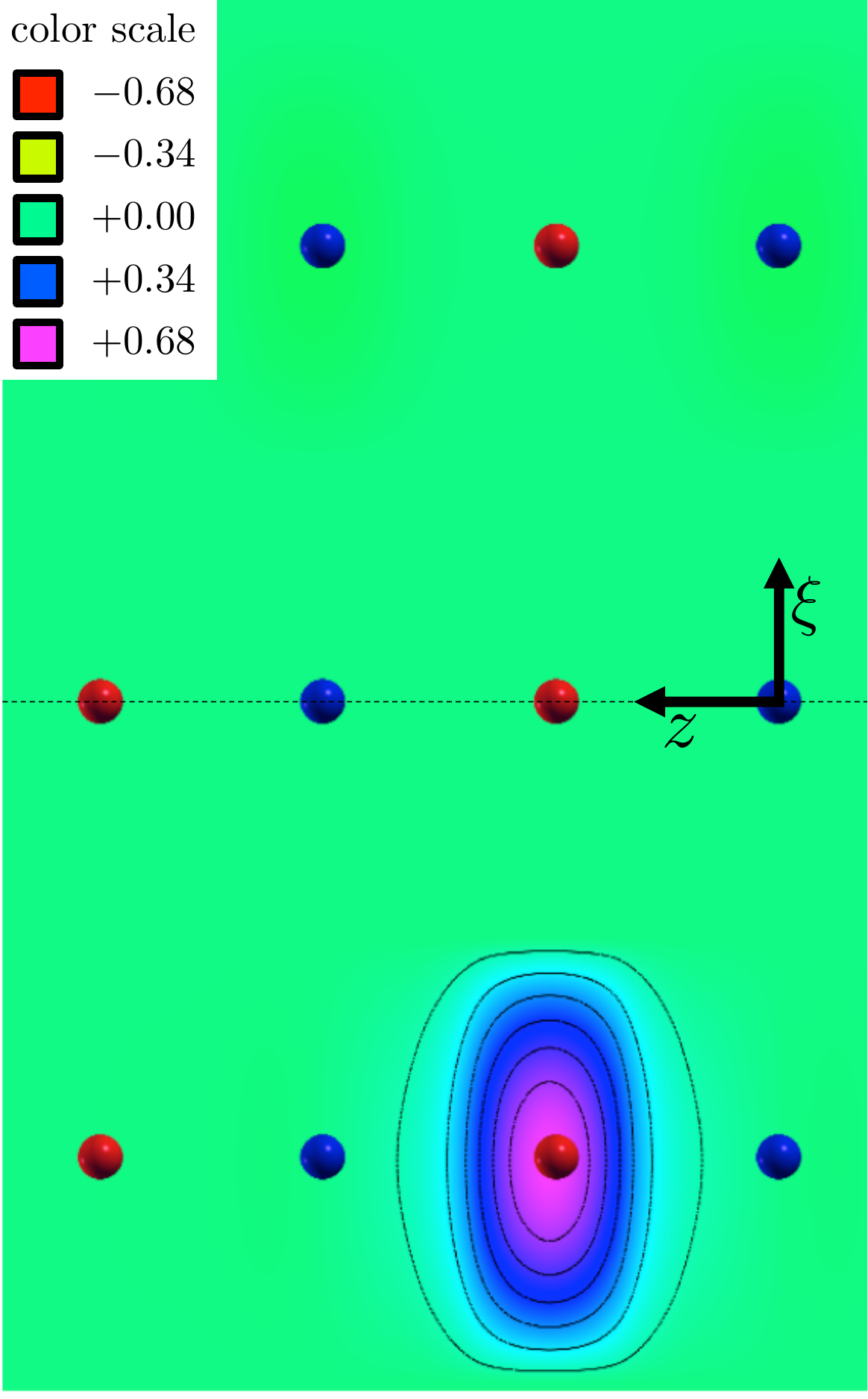}}\hfill
     \scalebox{0.35}{\includegraphics{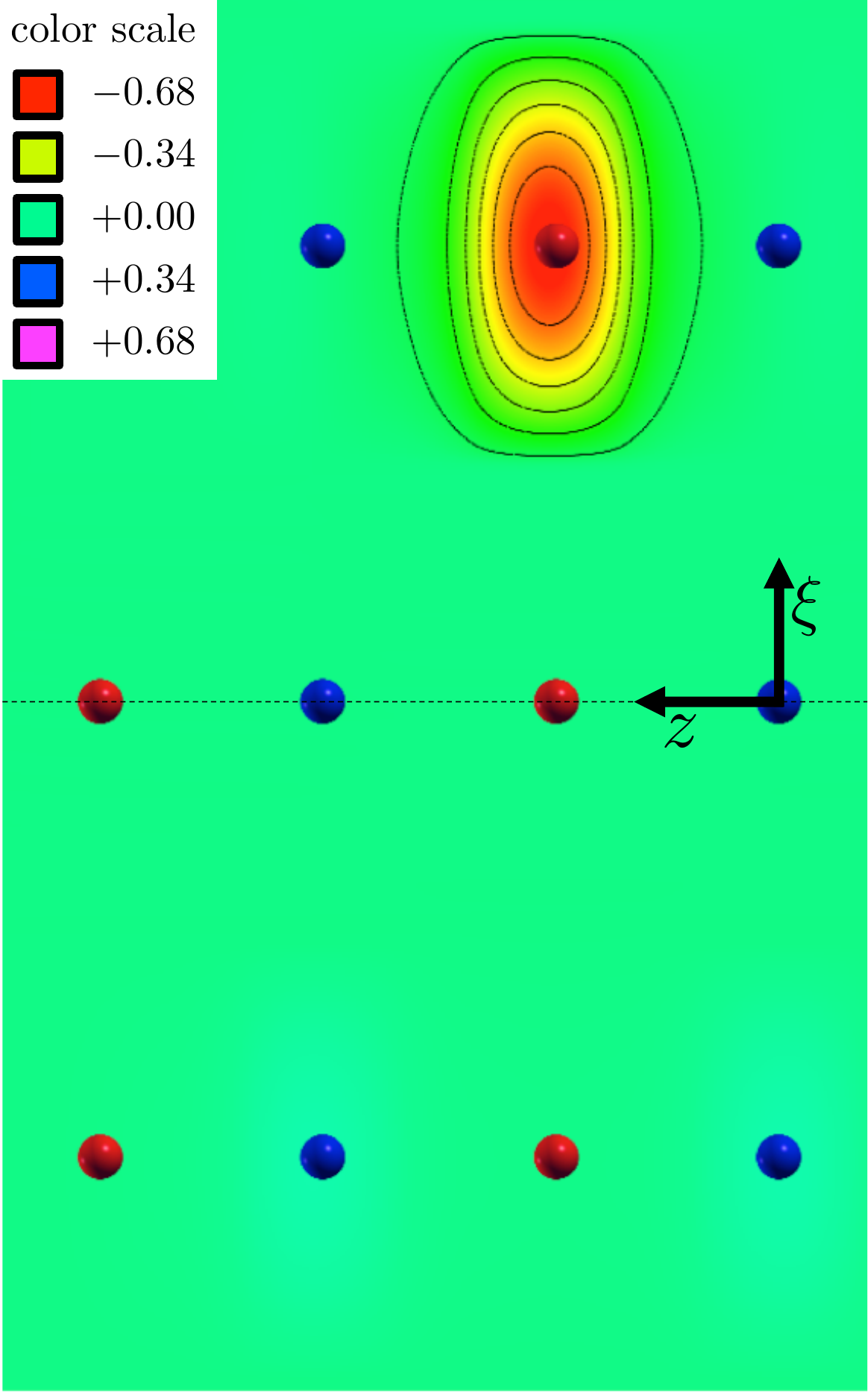}}
 \caption{Real-space distribution of the up (left) and down (right) component of the HDWF associated with spin moment rotations. Red balls refer to those moments $\hat{\vn n}_2$ which rotate with $\lambda$, and blue balls represent the positions of the fixed moments $\hat{\vn n}_1$. Black lines indicate contours of constant function value. The functions were plotted using the program XCrySDen (Ref.~\onlinecite{Kokalj2003}).}
 \label{fig:hdwf_stt}
\end{figure}

As shown in Fig.~\ref{fig:disp_stt}, \gwfshort{} succeed in the precise interpolation of the band structure of the family of Hamiltonians Eq.~\eqref{eq:stt_h} throughout the composite BZ of $k$ and $\lambda$. For this model, we also present in Fig.~\ref{fig:hdwf_stt} the localized real-space distribution of one of the low-energy \gwfshort{}. Remarkably, the spinor-valued HDWF turns out to be purely real. The individual components are both centered in the potential wells at $s=a/2$, where the exchange field is rotated. However, the down component is displaced by $+2$ lattice constants along the $\xi$ direction compared to the up component.

In the following, we give a simple argument for the shift in the coordinate $\xi$ between up and down components of the HDWF shown in Fig.~\ref{fig:hdwf_stt}. If we consider the deep-well limit $V_0\rightarrow\infty$, the two atoms in the unit cell do not hybridize and decouple completely. Thus, the problem is equivalent to finding the lowest-energy solution for a chain where all moments rotate with $\lambda$. The spin part $(\E{\I\lambda/2},-\E{-\I\lambda/2})$ of such a solution describes the rotation around the $z$ axis of a spin pointing initially in the $-x$ direction. We can perform a Fourier transformation of this spin part:
\begin{equation}
\begin{split}
 \begin{pmatrix}\mathcal W^\uparrow(\xi)\\ \mathcal W^\downarrow(\xi) \end{pmatrix}&= \sum\limits_\lambda \begin{pmatrix}\E{\I\frac{\lambda}{2}}\\-\E{-\I\frac{\lambda}{2}}
                                  \end{pmatrix} \E{\I\frac{\lambda}{2}\xi} \E{\I\phi_\lambda}\\ &=\sum\limits_\lambda \begin{pmatrix}\E{\I\lambda\left(\frac{\xi}{2}+\frac{1}{2}\right)+\I\phi_\lambda}\\-\E{\I\lambda\left(\frac{\xi}{2}-\frac{1}{2}\right)+\I\phi_\lambda}
                                  \end{pmatrix} \, ,
\end{split}
\end{equation}
where the phases $\E{\I\frac{\lambda}{2}\xi}$ guarantee the orthogonality, and the gauge freedom is represented by $\phi_\lambda$. It follows that $\mathcal W^\uparrow(\xi-2)=-\mathcal W^\downarrow(\xi)$. Consequently, the components of the spinor are opposite in sign and additionally shifted by two lattice constants in $\xi$-space.

\subsection{Mixed Berry curvature}
Recently, the mixed Berry curvature in $(\vn k,\hat{\vn m})$-space has been found to be important for spin-orbit torques, for the \DM{} interaction and for the charge of skyrmions \cite{Freimuth2014a,Freimuth2014,Kurebayashi2014,Freimuth2013}. This mixed Berry curvature is given by
\begin{equation}
 \Omega_{ij}^n(\vn k,\hat{\vn m}) = -2 \uvec_i \cdot \Bigg( \hat{\vn m}\times \Im\left\langle\frac{\partial u^{\text{gl}}_{\vn k \hat{\vn m} n}}{\partial \hat{\vn m}} \bigg| \frac{\partial u^{\text{gl}}_{\vn k \hat{\vn m} n}}{\partial k_j}\right\rangle \Bigg) \, ,
\label{eq:intrinsic_SOT}
\end{equation}
where $i$ and $j$ are Cartesian directions and $\uvec_i$ is the unit vector in the $i$-th Cartesian direction. If an electric field $\vn E$ is applied to a ferromagnet with broken inversion symmetry, the torque
\begin{equation}
 \vn T = -\frac{1}{N_{\vn k}} \sum\limits_{\vn k n}\sum\limits_{ij} e \Omega_{ij}^n(\vn k,\hat{\vn m}) \uvec_i E_j \, ,
\label{eq:torque}
\end{equation}
acts on the magnetization due to the mixed Berry curvature $\Omega_{ij}^n(\vn k,\hat{\vn m})$ \cite{Freimuth2014,Freimuth2014a}. Using spherical coordinates to express the magnetization direction such that $\hat{\vn m}=(\sin\theta\cos\phi,\sin\theta\sin\phi,\cos\theta)$, we can rewrite Eq.~\eqref{eq:intrinsic_SOT} as
\begin{equation}
\begin{split}
\Omega_{ij}^n(\vn k,\theta,\phi) = -2 \uvec_i \cdot \Bigg( &\uvec_{\phi} \Im \left\langle\frac{\partial u^{\text{gl}}_{\vn k \theta\phi n}}{\partial \theta} \bigg| \frac{\partial u^{\text{gl}}_{\vn k \theta\phi n}}{\partial k_j}\right\rangle \\
-&\uvec_{\theta} \frac{1}{\sin\theta}\Im \left\langle\frac{\partial u^{\text{gl}}_{\vn k \theta\phi n}}{\partial \phi} \bigg| \frac{\partial u^{\text{gl}}_{\vn k \theta\phi n}}{\partial k_j}\right\rangle \Bigg)\, .
\end{split}
\label{eq:mixedBC}
\end{equation}
If we construct \gwfshort{} for the Hamiltonian $H^{(\vn k,\theta,\phi)}$, we can use the generalized Wannier interpolation in order to evaluate Eq.~\eqref{eq:mixedBC}. In Sec.~\ref{sec:mn_chain_mae}, we demonstrated that \gwfshort{} can be constructed for $H^{(\vn k,\theta)}$. It is straightforward to extend the scheme of Sec.~\ref{sec:mn_chain_mae} to allow for variation of both $\theta$ and $\phi$. How to obtain the derivative $|\partial u_{\vn k n}/\partial k_j\rangle$ from the standard MLWF interpolation is discussed in detail in Ref.~\onlinecite{Wang2006}. The above derivatives $|\partial u^{\text{gl}}_{\vn k\theta\phi n}/\partial k_j\rangle$, $|\partial u^{\text{gl}}_{\vn k\theta\phi n}/\partial \theta\rangle$, and $|\partial u^{\text{gl}}_{\vn k\theta\phi n}/\partial \phi\rangle$ are calculated in the HDWF-interpolation scheme in a similar way. We suppress the superscript ``gl'' in the following.

The periodic parts of the Bloch-like functions are obtained from the \gwfshort{} by Fourier transformation:
\begin{equation}
\begin{split}
 |\tilde \varphi^{\phantom{\text{gl}}}_{\vn k \theta\phi n}\rangle &= |\tilde u^{\phantom{\text{gl}}}_{\vn k \theta\phi n} \artu^{\phantom{\text{gl}}}_{\theta\phi}\rangle \\
 &= \sum\limits_{\vn R}\sum\limits_{\Xi_\theta \Xi_\phi} \E{-\I \vn k \cdot(\vn r - \vn R)} \E{-\I \theta(\xi_\theta-\Xi_\theta)} \\
&\qquad\quad \ \, \, \, \times\E{-\I \phi(\xi_\phi-\Xi_\phi)} |\gwann_{\vn R \Xi_\theta \Xi_\phi n}\rangle \, .
\end{split}
\label{eq:u_hdwf}
\end{equation}
In order to acquire the periodic parts of the eigenfunctions of $H^{(\vn k,\theta,\phi)}$ (cf. Eq.~\eqref{eq:ham_interpol}), we need to apply an additional unitary matrix $V^{(\vn k,\theta,\phi)}$ (cf. Eq.~\eqref{eq:gen_WF_diag}):
\begin{equation}
 |\varphi^{\phantom{\text{gl}}}_{\vn k \theta\phi n}\rangle = \sum\limits_m |\tilde \varphi^{\phantom{\text{gl}}}_{\vn k \theta\phi m}\rangle V_{mn}^{(\vn k,\theta,\phi)} = | u^{\phantom{\text{gl}}}_{\vn k \theta\phi n} \artu^{\phantom{\text{gl}}}_{\theta\phi}\rangle\, ,
\end{equation}
where $|u^{\phantom{\text{gl}}}_{\vn k \theta\phi n}\rangle = \sum_m | \tilde u^{\phantom{\text{gl}}}_{\vn k \theta\phi m} \rangle V_{mn}^{(\vn k,\theta,\phi)}$. Accordingly, the $\theta$-derivative of the latter is given by
\begin{equation}
\begin{split}
 \bigg|\frac{\partial u^{\phantom{\text{gl}}}_{\vn k \theta \phi n}}{\partial \theta}\bigg\rangle = &\sum\limits_m \bigg|\frac{\partial \tilde u^{\phantom{\text{gl}}}_{\vn k \theta \phi m}}{\partial \theta}\bigg\rangle V_{mn}^{(\vn k,\theta,\phi)} \\
+ &\sum\limits_m | u^{\phantom{\text{gl}}}_{\vn k \theta\phi m}\rangle D_{mn}^{(\vn k,\theta,\phi)}\, ,
\end{split}
\label{eq:theta_derivative}
\end{equation}
where the derivative of the unitary transformation is written as $\partial V^{(\vn k,\theta,\phi)}/\partial \theta = V^{(\vn k,\theta,\phi)}D^{(\vn k,\theta,\phi)}$. Introducing the abbreviations $\gpara=(\theta,\phi)$, $\gpos=(\xi_\theta,\xi_\phi)$, and $\gdlatt=(\Xi_\theta,\Xi_\phi)$, we can use for the first term that
\begin{equation}
 \begin{split}
  &\bigg\langle \tilde u^{\phantom{\text{gl}}}_{\vn k \theta\phi n} \bigg| \frac{\partial \tilde u^{\phantom{\text{gl}}}_{\vn k \theta\phi m}}{\partial \theta}\bigg\rangle 
=\sum\limits_{\vn R \gdlatt}\sum\limits_{\vn R^\prime \gdlatt^\prime} \E{\I \vn k \cdot(\vn R-\vn R^\prime)} \E{\I\gpara\cdot(\gdlatt-\gdlatt^\prime)} \\
&\qquad\qquad\qquad\qquad\,\, \,  \times  \langle \gwann_{\vn R^\prime \gdlatt^\prime n} | \left[-\I(\xi_\theta-\Xi_\theta)\right]| \gwann_{\vn R\gdlatt m}\rangle \\
&=-\I\sum\limits_{\vn R \gdlatt}\sum\limits_{\vn R^\prime \gdlatt^\prime}  \E{\I \vn k \cdot(\vn R-\vn R^\prime)} \E{\I\gpara\cdot(\gdlatt-\gdlatt^\prime)}\langle \gwann_{\vn R^\prime \gdlatt^\prime n} |\xi_\theta | \gwann_{\vn R\gdlatt m}\rangle \, ,
 \end{split}
 \label{eq:connection_interpolation}
\end{equation}
which follows from Eq.~\eqref{eq:u_hdwf} and $\langle \artu^{\phantom{\text{gl}}}_{\gpara}|\nabla_{\gpara}\artu^{\phantom{\text{gl}}}_{\gpara}\rangle=0$. The matrix elements $\langle \gwann_{\vn R^\prime \gdlatt^\prime n} |\xi_\theta | \gwann_{\vn R\gdlatt m}\rangle$ can be computed by generalizing Eq.~\eqref{eq:gen_WF_center_xi}:
\begin{equation}
\begin{split}
 \langle \gwann_{\vn R^\prime \gdlatt^\prime n} | \gpos | \gwann_{\vn R \gdlatt m}\rangle = \frac{\I}{N_{\vn k}N_{\gpara}} &\sum\limits_{\vn k \gpara} \E{-\I \vn k \cdot(\vn R-\vn R^\prime)} \E{-\I \gpara\cdot(\gdlatt-\gdlatt^\prime)} \\
&\ \, \times\langle \tilde u^{\phantom{\text{gl}}}_{\vn k \gpara n} | \nabla_{\gpara} | \tilde u^{\phantom{\text{gl}}}_{\vn k \gpara m}\rangle \, .
\end{split}
\end{equation}
Similar off-diagonal matrix elements are available in the \wprog{} code for the case of standard MLWFs. They are obtained by approximating the gradient by finite differences \cite{Marzari1997}:
\begin{equation}
 \langle W_{\vn 0 n} | \vn r | W_{\vn R m}\rangle = \frac{\I}{N_{\vn k}} \sum\limits_{\vn k \vn b} \E{-\I \vn k \cdot \vn R} w_b \vn b \left(M_{nm}^{(\vn k,\vn b)} - \delta_{nm}\right) \, .
\label{eq:std_fd}
\end{equation}
It is straightforward to generalize Eq.~\eqref{eq:std_fd} for the HDWF case based on the overlaps in Eq.~\eqref{eq:gen_WF_M_k} and Eq.~\eqref{eq:gen_WF_M_q}. For the second term in Eq.~\eqref{eq:theta_derivative} we can use \cite{Wang2006}
\begin{equation}
  D_{mn}^{(\vn k,\theta,\phi)} = \begin{dcases} \frac{\left\langle \varphi^{\phantom{\text{gl}}}_{\vn k \theta\phi m} \left| \frac{\partial H^{(\vn k,\theta,\phi)}}{\partial \theta} \right| \varphi^{\phantom{\text{gl}}}_{\vn k \theta\phi n}\right\rangle}{\mathcal{E}_{\vn k \theta\phi n}-\mathcal{E}_{\vn k \theta\phi m}} & \text{if } n\neq m \\ 0 & \text{if } n=m \end{dcases}
\end{equation}
which can be evaluated for any $(\vn k,\theta,\phi)$ from generalized Wannier interpolation. Analogously, the derivatives $|\partial u^{\phantom{\text{gl}}}_{\vn k\theta\phi n}/\partial k_j\rangle$ and $|\partial u^{\phantom{\text{gl}}}_{\vn k\theta\phi n}/\partial \phi\rangle$ are constructed within the formalism of \gwfshort{}.

\section{Summary}
\label{sec:conclusions}
We introduce the concept and formalism of higher-dimensional Wannier functions (\gwfshort{}) to describe the electronic structure of multi-parameter Hamiltonians $H^{(\vn k,\gpara)}$, where $\gpara$ is an external periodic parameter. The introduction of an auxiliary space $\gpos$ solves the fundamental problem of non-orthogonality of usual Bloch states in such a situation. Analogously to maximally-localized Wannier
functions, we define HDWFs as Fourier transformations of higher dimensional product states carrying a dependence on $\vn k$ and $\gpara$. A minimal and accurate interpolation of multi-parameter Hamiltonians is established using \gwfshort{}. The implementation of the necessary machinery for the construction of \gwfshort{} from \abinitio{} within the FLAPW method is discussed. In order to achieve a maximal localization in the extended space of $\vn r$ and $\gpos$, we adapt the \wprog{} program.

The application of the formalism to a one-dimensional Mn chain with the spin-spiral vector as an external parameter reveals an excellent agreement with direct \firstprinciples{} calculations, and enables the simplified extraction of Heisenberg exchange constants and spin stiffness. Treating the direction of the ferromagnetic magnetization in real space as an external parameter, we are able to apply the HDWFs machinery to compute the magneto-crystalline anisotropy energy of the Mn chain. Although the corresponding energy scale is very small and thus more difficult to capture, \gwfshort{} interpolate accurately the energy $E(\theta)$ as a function of the magnetization direction. We outline various physical problems to which \gwfshort{} could be applied efficiently, e.g., disorder treated within VCA. We emphasize further the advantages associated with the evaluation of linear response coefficients such as AHE and spin torques. A formula for the generalized interpolation of the ferroelectric polarization along any insulating $\gpara$-path is provided. Finally, \gwfshort{} could prove useful in the topological characterization of complex multi-parameter systems as they allow for the simplified evaluation of mixed Berry curvatures.

\section*{Acknowledgments}
We gratefully acknowledge computing time on the supercomputers JUQUEEN and JUROPA at
J\"ulich Supercomputing Center as well as at the JARA-HPC cluster of RWTH Aachen, and funding under the HGF-YIG programme VH-NG-513 and SPP 1538 of DFG.

\appendix

\section{Details on the FLAPW implementation}
\label{app:details_FLAPW}
Expressions for the overlaps $M_{mn}^{(\vn k,\vn b)}$ between periodic parts of the Bloch states at neighboring crystal momenta and projections $A_{mn}^{(\vn k)}$ were already derived for an implementation of MLWFs within FLAPW \cite{Freimuth2008}. The evaluation of Eq.~\eqref{eq:gen_WF_M_q} requires additionally the construction of the overlaps
\begin{equation}
 \mathcal{M}_{mn}^{(\vn q,\vn b)}(\vn k) = \sum\limits_\sigma \langle u^\sigma_{\vn k \vn q m} | u^\sigma_{\vn k\,\vn q+\vn b\,n}\rangle 
 \label{eq:fleur_overlap}
\end{equation}
with the vector $\vn b=b \hat{\vn b}$ connecting the two $\vn q$-points, and $\sigma=\uparrow,\downarrow$. Because of the real-space partition into muffin tin spheres (MT) and the interstitial region (INT), these matrix elements decompose:
\begin{equation}
 \mathcal{M}_{mn}^{(\vn q,\vn b)}(\vn k) = \left.\mathcal{M}_{mn}^{(\vn q,\vn b)}(\vn k)\right|_{\text{INT}} + \sum\limits_\mu \left.\mathcal{M}_{mn}^{(\vn q,\vn b)}(\vn k)\right|_{\text{MT}_\mu} \, .
\end{equation}
Here, $\mu$ labels the different atoms in the unit cell. Further contributions arise in film calculations (cf. Appendix~\ref{app:film}), the study of one-dimensional geometries (cf. Appendix~\ref{app:od}), and when the FLAPW basis set is supplemented with local orbitals. Due to the generalized Bloch theorem, Eq.~\eqref{eq:gen_bloch_theorem}, the overlaps in Eq.~\eqref{eq:fleur_overlap} can be rewritten:
\begin{equation}
 \mathcal{M}_{mn}^{(\vn q,\vn b)}(\vn k) = \sum\limits_\sigma \int \E{\pm \I \frac{\vn b}{2}\cdot\vn r} \left(\Psi^\sigma_{\vn k \vn q m}(\vn r)\right)^* \Psi^\sigma_{\vn k\,\left[\vn q+\vn b\right]\,n}(\vn r)\,\D\vn r \, ,
\label{eq:fleur_Mmn}
\end{equation}
where the upper (lower) sign is associated with the up-spin (down-spin) of the Bloch states. The expression $\left[\vn q\right]$ refers to a backfolding of the momentum $\vn q$ into the first BZ by the subtraction of a reciprocal lattice vector $\vn G(\vn q)$, namely, $\left[\vn q\right] = \vn q - \vn G(\vn q)$. As a consequence of the doubled BZ of $\vn q$-points, $\vn G(\vn q)$ is twice as large as a usual reciprocal lattice vector (see also remark below Eq.~\eqref{eq:gen_bloch_theorem}).

Within the muffin tin sphere centered around the $\mu$-th atom, which is located at the position $\vn{\tau}_\mu$, plane-waves do not succeed in describing the physics in presence of the singular atomic potential. Thus, the Bloch states are expanded in terms of radial solutions $u_l$ of the scalar relativistic equation at band-averaged energies, related derivatives with respect to energy $\dot{u}_l$, and the spherical harmonics $Y_L$ where $L=(l,l_z)$ represents the set of angular momentum quantum numbers. Accordingly, the single-particle wave function is given by
\begin{equation}
\begin{split}
 \Psi^\sigma_{\vn k \vn q n} (\vn r) = \sum\limits_{L}\big[ &a^{\mu,\sigma}_{Ln}(\vn k,\vn q) u^{\mu,\sigma}_l(r_{\mu}) \\
+ &b^{\mu,\sigma}_{Ln}(\vn k, \vn q)\dot{u}^{\mu,\sigma}_l(r_{\mu})\big] Y_L(\hat{\vn r}_{\mu}) \, .
\end{split}
 \label{eq:fleur_mt_psi}
\end{equation}
Here, $a^{\mu,\sigma}_{Ln}$ and $b^{\mu,\sigma}_{Ln}$ are expansion coefficients in the $\mu$-th muffin tin and the position relative to the nucleus is denoted as $\vn r_{\mu}=\vn r-\vn{\tau}_\mu$. If we employ the Rayleigh expansion
\begin{equation} 
 \E{\mp \I\vn b\cdot\vn r} = 4\pi \E{\mp \I\vn b\cdot\vn\tau_\mu} \sum\limits_{L} (\mp 1)^l \I^l j_l(r_{\mu} b) Y_L(\hat{\vn b}) \left(Y_L(\hat{\vn r}_{\mu})\right)^*
\end{equation}
of the plane-wave factor in Eq.~\eqref{eq:fleur_Mmn} into spherical harmonics, the muffin tin contribution to the overlaps between periodic parts assumes the form
\begin{equation}
\begin{split}
 \mathcal{M}_{mn}^{(\vn q,\vn b)}&(\vn k)\left.\vphantom{M_{mn}^{(\vn q,\vn b)}}\right|_{\text{MT}_\mu} = 4 \pi \sum\limits_\sigma\E{\pm \I \frac{\vn b}{2}\cdot\vn \tau_\mu}\\
\times&\sum\limits_{LL^\prime} \left[ \left(a^{\mu,\sigma}_{Lm}(\vn k,\vn q)\right)^* a^{\mu,\sigma}_{L^\prime n}(\vn k,\left[\vn q+\vn b\right]) t^{\mu,LL^\prime}_{11} (\vn b,\sigma)\right. \\
 +&\left. \left(a^{\mu,\sigma}_{Lm}(\vn k,\vn q)\right)^* b^{\mu,\sigma}_{L^\prime n}(\vn k,\left[\vn q+\vn b\right]) t^{\mu,LL^\prime}_{12} (\vn b,\sigma)\right.\\
 +&\left.\left(b^{\mu,\sigma}_{Lm}(\vn k,\vn q)\right)^* a^{\mu,\sigma}_{L^\prime n}(\vn k,\left[\vn q+\vn b\right]) t^{\mu,LL^\prime}_{21} (\vn b,\sigma)\right. \\
+&\left.\left(b^{\mu,\sigma}_{Lm}(\vn k,\vn q)\right)^* b^{\mu,\sigma}_{L^\prime n}(\vn k,\left[\vn q+\vn b\right]) t^{\mu,LL^\prime}_{22} (\vn b,\sigma)\right] \, .
 \end{split}
 \label{eq:fleur_sph_mmn}
\end{equation}
Here, the radial solutions, their energy derivatives, and the spherical Bessel functions $j_l$ enter through the $t$-coefficients defined as
\begin{align}
\begin{split}
 t^{\mu,L''L}_{11} &(\vn b,\sigma) = \sum_{L^\prime} \mathcal{G}_{LL^\prime L''}(\hat{\vn b}) \\
&\times \int r_{\mu}^2 \,j_{l^\prime}\left(\frac{r_{\mu} b}{2}\right) u^{\mu,\sigma}_l(r_{\mu}) u^{\mu,\sigma}_{l''}(r_{\mu})\,\D r_{\mu}\, ,
\label{eq:t_integral_spinspiral}
\end{split} \\
\begin{split}
 t^{\mu,L''L}_{12} &(\vn b,\sigma) = \sum_{L^\prime} \mathcal{G}_{LL^\prime L''}(\hat{\vn b}) \\
&\times \int r_{\mu}^2 \,j_{l^\prime}\left(\frac{r_{\mu}b}{2}\right) \dot{u}^{\mu,\sigma}_l(r_{\mu}) u^{\mu,\sigma}_{l''}(r_{\mu})\,\D r_{\mu} \, ,
\end{split}
\end{align}
and likewise for $t_{21}$ and $t_{22}$. If we choose a uniform Monkhorst-Pack grid to sample the BZ of spin-spiral parameters, the above integrals become independent of the $\vn q$-point such that they may be calculated once and for all at the very beginning. The abbreviation
\begin{equation}
\mathcal{G}_{LL^\prime L''}(\hat{\vn b}) = \I^{l^\prime} (\pm 1)^{l^\prime} Y_{L^\prime}(\hat{\vn b}) G_{LL^\prime L''}
\label{eq:fleur_G_temp}
\end{equation}
incorporates the Gaunt coefficients $G_{LL^\prime L''}$, which are given by
\begin{equation}
 G_{LL^\prime L''} = \int Y_{L}(\hat{\vn r}_{\mu}) \left(Y_{L^\prime}(\hat{\vn r}_{\mu})\right)^*\left(Y_{L''}(\hat{\vn r}_{\mu})\right)^*\,\D\Omega \, .
\end{equation}
The expressions above are easily extended when local orbitals are employed in the basis set.

In FLAPW, the Bloch states are expanded using plane-waves with reciprocal lattice vectors $\vn G$ in the interstitial region. Thus, the wave function assumes a form in line with the generalized Bloch theorem:
\begin{equation}
 \Psi^\sigma_{\vn k \vn q n} (\vn r) = \frac{1}{\sqrt{V}} \sum\limits_{\vn G} c^\sigma_{\vn G}(\vn k,\vn q,n) \E{\I \left(\vn k \mp \frac{\vn q}{2} + \vn G\right)\cdot\vn r} \, .
 \label{eq:fleur_int_psi}
\end{equation}
Defining the Fourier transformation of the step function $\Theta_{\text{INT}}$ cutting out the interstitial region by
\begin{equation}
 \Theta_{\vn G} = \frac{1}{V}\int_{\text{INT}} \E{-\I \vn G\cdot \vn r}\,\D\vn r = \frac{1}{V}\int\limits\E{-\I \vn G\cdot\vn r}\, \Theta_\text{INT}(\vn r)\,\D\vn r \, ,
\label{eq:FT_theta_INT}
\end{equation}
we can write the interstitial contribution to the overlap elements of the periodic parts at neighboring spin-spiral parameters, Eq.~\eqref{eq:fleur_Mmn}, as
\begin{equation}
\begin{split}
\left.\mathcal{M}_{mn}^{(\vn q,\vn b)}(\vn k)\right|_{\text{INT}}&= \sum\limits_{\vn G \vn G^\prime \sigma} \left(c^\sigma_{\vn G}(\vn k,\vn q,m)\right)^* c^\sigma_{\vn G^\prime}(\vn k,\left[\vn q+\vn b\right],n) \\
&\qquad\times\Theta_{\mp \frac{\vn G(\vn q+\vn b)}{2}+\vn G-\vn G^\prime} \, .
\end{split}
 \label{eq:fleur_int_mmn}
\end{equation}

The shapes of the overlaps Eq.~\eqref{eq:fleur_sph_mmn} and Eq.~\eqref{eq:fleur_int_mmn} differ slightly from those of the $M_{mn}^{(\vn k,\vn b)}$ contributions described in Ref.~\onlinecite{Freimuth2008}. First, the expansion coefficients carry a new dependence on the spin-spiral vector $\vn q$. An additional spin-dependent sign arises from the generalized Bloch theorem in Eqs.~\eqref{eq:fleur_sph_mmn}, \eqref{eq:fleur_G_temp}, and~\eqref{eq:fleur_int_mmn}. Finally, the vectors $\vn b$ and $\vn G(\vn q+\vn b)$ occur both with a factor of $1/2$ in Eqs.~\eqref{eq:fleur_sph_mmn}, \eqref{eq:fleur_int_mmn}, and the definition of the $t$-integrals.

To construct first-guess \gwfshort{}, the projections of the Bloch states onto localized trial orbitals $g_n$ have to be evaluated within FLAPW according to Eq.~\eqref{eq:gen_WF_A}. These trial orbitals are chosen to be zero everywhere except for the $\mu$-th muffin tin sphere to which the corresponding first-guess should be associated. The expansion coefficients in $g_n(\vn r) = \sum _L c_{Ln} \tilde u_{l}(r_{\mu}) Y_L(\hat{\vn r}_{\mu})$ control the angular character of the trial functions \cite{Freimuth2008}. The radial function $\tilde u_l$ can be chosen, for example, as the \firstprinciples{} solution $u^\mu_l$ to the radial Schr\"odinger equation. Then, projections $\langle \Psi_{\vn k \vn q m} | g_n \rangle$ are computed according to
\begin{equation}
\begin{split}
 &\sum\limits_{L\sigma} \left[\left(a^{\mu,\sigma}_{Lm}(\vn k,\vn q)\right)^*c^\sigma_{L n} \int r_{\mu}^2 \,u^{\mu,\sigma}_l(r_{\mu}) \tilde u^{\sigma}_{l}(r_{\mu})\,\D r_{\mu} \right.\\
&\quad  +\left. \left(b^{\mu,\sigma}_{Lm}(\vn k, \vn q)\right)^*c^\sigma_{L n} \int r_{\mu}^2 \,\dot{u}^{\mu,\sigma}_l(r_{\mu})\tilde u^{\sigma}_{l}(r_{\mu})\,\D r_{\mu}\,\right]
\end{split}
\end{equation}
if the orthogonality of the spherical harmonics is exploited. Except for the $\vn q$-dependence of the expansion coefficients, these expressions are similar to those described in Ref.~\onlinecite{Freimuth2008} for standard MLWFs.

\section{Vacuum contribution to the overlaps \texorpdfstring{$\mathcal M_{mn}^{(\vn q,\vn b)}(\vn k)$ }{}in film calculations}
 \label{app:film}
In the study of two-dimensional geometries using the film implementation of the \fleur{} program, an additional contribution to the matrix elements in Eq.~\eqref{eq:fleur_Mmn} occurs as a consequence of the presence of two semi-infinite vacua \cite{Krakauer1979}. The Bloch states in each of the vacua, which extend from $-\infty$ to $-\mathcal D/2$ as well as $\mathcal D/2$ to $\infty$, are represented by
\begin{equation}
 \Psi^\sigma_{\vn k_\parallel \vn q_\parallel n}(\vn r) = \sum\limits_{\vn G_{\parallel}} \psi^{n,\sigma}_{\vn G_\parallel}(\vn k_\parallel,\vn q_\parallel,z) \E{\I\left(\vn k_\parallel\mp\frac{\vn q_\parallel}{2}+\vn G_\parallel\right)\cdot\vn r} \, .
 \label{eq:fleur_film_psi}
\end{equation}
Here, $\vn k_\parallel$ and $\vn q_\parallel$ are both considered to lie within an according two-dimensional BZ associated with the film plane, which is supposed to be perpendicular to the $z$-axis. The function
\begin{equation}
\begin{split}
  \psi^{n,\sigma}_{\vn G_\parallel}(\vn k_\parallel, \vn q_{\parallel}, z) = &a^\sigma_{\vn G_\parallel n}(\vn k_\parallel,\vn q_\parallel) u^\sigma_{\vn G_\parallel}(\vn k_\parallel,\vn q_\parallel,z)\\
+ &\, b^\sigma_{\vn G_\parallel n}(\vn k_\parallel,\vn q_\parallel)\dot{u}^\sigma_{\vn G_\parallel}(\vn k_\parallel,\vn q_\parallel,z)
\label{eq:b2}
\end{split}
\end{equation}
includes the one-dimensional solutions of the Schr\"odinger equation in the corresponding vacuum region $u_{\vn G_\parallel}$ and their energy derivatives $\dot{u}_{\vn G_\parallel}$. For convenience, the abbreviations
\begin{equation}
\begin{split}
\beta_{\vn G_\parallel \vn G^\prime_\parallel}^{mn,\sigma}(\vn k_\parallel,\vn q_\parallel,[\vn q_\parallel+\vn b],z) =&  \left(\psi^{m,\sigma}_{\vn G_\parallel}(\vn k_\parallel,\vn q_\parallel,z)\right)^* \\
\times &\psi^{n,\sigma}_{\vn G^\prime_\parallel}(\vn k_\parallel,\left[\vn q_\parallel+\vn b\right],z)
\end{split}
\end{equation}
and $\mathcal{G}_\parallel = \vn G_\parallel - \vn G^\prime_\parallel \mp\vn G_\parallel(\vn q_\parallel+\vn b)/2$ are introduced. Consequently, the contribution of the vacuum extending from $\mathcal D/2$ to $\infty$ to the overlap matrix elements between periodic parts, Eq.~\eqref{eq:fleur_Mmn}, evaluates to
\begin{equation}
\begin{split}
\left.\mathcal{M}_{mn}^{(\vn q_\parallel,\vn b)}(\vn k_\parallel)\right|_{\text{FILM}}= &\sum\limits_\sigma\sum\limits_{\vn G_\parallel \vn G^\prime_\parallel} S_\parallel\, \delta_{\mathcal{G}_\parallel} \int_{\mathcal D/2}^{\infty} \E{\pm \I\frac{G_z(\vn q_\parallel+\vn b)}{2}z} \\
 \times &\beta_{\vn G_\parallel \vn G^\prime_\parallel}^{mn,\sigma}(\vn k_\parallel,\vn q_\parallel,[\vn q_\parallel+\vn b],z)\,\D z \, ,
\end{split}
\label{eq:b4}
\end{equation}
where the unit cell area with respect to the film plane is denoted as $S_{\parallel}$. The other contribution from the second vacuum region is derived analogously. Compared to the contribution to the usual overlaps $M_{mn}^{(\vn k,\vn b)}$, Ref.~\onlinecite{Freimuth2008}, the function $\psi^{n,\sigma}_{\vn G_\parallel}$ of Eq.~\eqref{eq:b2} carries a dependence on $\vn q_\parallel$. Additionally, the reciprocal lattice vector $\vn G(\vn q_\parallel+\vn b)$ occurs with a spin-dependent sign and a factor of $1/2$ in the definition of $\mathcal{G}_\parallel$ and Eq.~\eqref{eq:b4}.

\section{Vacuum contribution to the overlaps \texorpdfstring{$\mathcal M_{mn}^{(\vn q,\vn b)}(\vn k)$ }{}in one-dimensional calculations}
 \label{app:od}
The density functional theory code \fleur{} treats one-dimensional systems as cylinders with radius $R_{\text{vac}}$ embedded in surrounding vacuum \cite{Mokrousov2005}. The cylinder axis points along the $z$ direction. Using cylindrical coordinates in real space $\vn r = (z,r,\phi)$ and reciprocal space $\vn G = (G_z,G_r,G_\phi)$, we express the single-particle wave function in the vacuum as
\begin{equation}
 \Psi^\sigma_{k_z q_z n}(\vn r) = \sum\limits_{P} \psi^{n,\sigma}_{P}(k_z, q_z,r) \E{\I p\phi} \E{\I\left(k_z\mp\frac{q_z}{2}+G_z\right)z} \, ,
\label{eq:fleur_od_psi}
\end{equation}
where $k_z$ as well as $q_z$ are drawn from a one-dimensional BZ, and the integer $p$ labels the cylindrical angular harmonics. The variable $P$ denotes the set of $(G_z,p)$ with respect to which the summation is performed. Radial solutions $u_{P}$ to the Schr\"odinger equation in the vacuum region and related energy derivatives $\dot{u}_P$ enter the expression through
\begin{equation}
\begin{split}
 \psi^{n,\sigma}_{P}(k_z, q_z, r) =& a^{n,\sigma}_{P}(k_z, q_z) u^\sigma_{P}(k_z, q_z,r) \\
+& b^{n,\sigma}_{P}(k_z, q_z)\dot{u}^\sigma_{P}(k_z, q_z,r) \, .
\end{split}
\end{equation}
For convenience, the abbreviations
\begin{equation}
\begin{split}
\beta_{PP^\prime}^{mn,\sigma}(k_z,q_z,[q_z+\vn b],r) =& \left(\psi^{m,\sigma}_{P}(k_z,q_z,r)\right)^*\\ \times &\psi^{n,\sigma}_{P^\prime}(k_z,\left[q_z+\vn b\right],r)
\end{split}
\end{equation}
and $\mathcal{G}_z = G_z - G^\prime_z \mp G_z(q_z+\vn b)/2$ are introduced such that the corresponding overlap elements, Eq.~\eqref{eq:fleur_Mmn}, associated with the presence of the vacuum assume the form
\begin{equation}
\begin{split}
 \left.\mathcal{M}_{mn}^{(q_z,\vn b)}(k_z)\right|_{\text{OD}}=&\sum\limits_\sigma \sum\limits_{P P^\prime} \int_{\text{VAC}} \beta_{PP^\prime}^{mn,\sigma}(k_z,q_z,[q_z+\vn b],r) \\
\times& \E{-\I\mathcal{G}_z z}\,\E{\pm \I\frac{\vn G_\parallel(q_z+\vn b)}{2}\cdot\vn r_\parallel}\,\E{\I(p^\prime - p)\phi}\,\D\vn r \, .
\end{split}
\end{equation}
Here, $\vn r_\parallel$ shall refer to the $x$- and $y$-component of the real-space vector $\vn r = (\vn r_\parallel, z)$ and similar for the reciprocal lattice vector $\vn G_\parallel(q_z+\vn b)$, which shifts the momentum back into the first BZ. Exploiting then the plane-wave expansion into cylindrical coordinates
\begin{equation}
  \E{\mp \I\vn G\cdot \vn r} = \E{\mp \I G_z z} \sum\limits_p \I^p (\mp 1)^p \E{\mp \I p\left(\phi-\phi_{\vn G}\right)} J_p(G_r r) \, ,
\end{equation}
we arrive finally at the vacuum contribution to the overlaps of periodic parts at neighboring $\vn q$, Eq.~\eqref{eq:fleur_Mmn}, in case of one-dimensional calculations:
\begin{equation}
\begin{split}
 \left.\mathcal{M}_{mn}^{(q_z,\vn b)}(k_z)\right|_{\text{OD}}=& \sum\limits_\sigma \sum\limits_{P P^\prime}  (\mp 1)^{p^\prime-p} \,\I^{p-p^\prime} \E{\I(p^\prime-p)\phi_{\vn G(q_z+\vn b)}}\\
\times & \ell\,\delta_{\mathcal{G}_z} \int_{R_\text{vac}}^{\infty} r J_{p^\prime-p}\left(\frac{G_r(q_z+\vn b) r}{2}\right)\\
\times & \beta_{PP^\prime}^{mn,\sigma}(k_z,q_z,[q_z+\vn b],r) \, \D r \, .
\end{split}
\label{eq:c6}
\end{equation}
Here, $J_p$ represents the cylindrical Bessel function of order $p$, and $\ell=2\pi T$ with the lattice constant $T$ along the axis of translational invariance. In contrast to the implementation of the usual overlaps $M_{mn}^{(\vn k,\vn b)}$, Ref.~\onlinecite{Freimuth2008}, a spin-dependent sign arises from the generalized Bloch theorem in Eq.~\eqref{eq:c6}. The lattice vector $\vn G(q_z+\vn b)$ occurs further with an additional factor $1/2$ in the argument of the cylindrical Bessel function, and the definition of $\mathcal{G}_z$.

\section{Calculation of \texorpdfstring{$\mathcal M_{mn}^{(\theta,b)}(\vn k)$}{Mmn} within FLAPW}
\label{app:theta_FLAPW}
Knowledge of the overlaps between periodic parts of the Bloch states at neighboring angles $\theta$ and $\theta+b$ is required to construct \gwfshort{} when magnetization direction plays the role of the additional external parameter. Within the second-variation scheme \cite{Li1990} used in this work, the spin quantization axis of the wave functions is identical to the magnetization direction, which we characterize by an angle $\theta$. Using the rotation
\begin{equation}
 \chi(\theta) = \begin{pmatrix} \cos\frac{\theta}{2} & -\sin\frac{\theta}{2} \\[6pt] \sin\frac{\theta}{2} & \cos\frac{\theta}{2} \end{pmatrix}\, ,
\end{equation}
we transform therefore all wave functions to the very same global frame in order to evaluate the overlaps
\begin{equation}
\begin{split}
\mathcal{M}_{mn}^{(\theta,b)}(\vn k) &= \sum\limits_\sigma \langle u^{\sigma,\text{gl}}_{\vn k \theta m} | u^{\sigma,\text{gl}}_{\vn k\, \theta+b\, n}\rangle \\
&= \sum\limits_{\sigma\sigma^\prime} \left[\chi^\dagger(\theta) \chi(\theta+b)\right]_{\sigma\sigma^\prime} \langle u^{\sigma}_{\vn k \theta m} | u^{\sigma^\prime}_{\vn k\, \theta+b\, n}\rangle \, .
\label{eq:Mmn_theta}
\end{split}
\end{equation}
Here, the periodic part $u_{\vn k \theta n}$ in the local coordinate frame was transformed to the global one by $u^{\text{gl}}_{\vn k \theta n}=\chi(\theta)u_{\vn k \theta n}$, and $\sigma = \uparrow,\downarrow$. Keeping in mind Eq.~\eqref{eq:Mmn_theta}, we present in the following the necessary FLAPW expressions for the calculation of the overlaps in the local spin frame.

The standard expansion of the wave function into plane waves is used in the interstitial region with the expansion coefficients carrying now a dependence on the angle $\theta$:
\begin{equation}
 \Psi^\sigma_{\vn k \theta n} (\vn r) = \frac{1}{\sqrt{V}} \sum\limits_{\vn G} c^\sigma_{\vn G}(\vn k,\theta,n) \E{\I \left(\vn k + \vn G\right)\cdot\vn r} \, .
 \label{eq:fleur_int_psi_soc}
\end{equation}
Thus, the overlaps of lattice periodic parts in the local coordinate frame, Eq.~\eqref{eq:chain_temp_beta}, assume the form 
\begin{equation}
\begin{split}
\langle u^{\sigma}_{\vn k \theta m} &| u^{\sigma^\prime}_{\vn k\, \theta+b\, n}\rangle\Big|_{\text{INT}} =\\
&=\sum\limits_{\vn G \vn G^\prime} \left(c^\sigma_{\vn G}(\vn k,\theta,m)\right)^* c^{\sigma^\prime}_{\vn G^\prime}(\vn k,\left[\theta+b\right],n) \Theta_{\vn G-\vn G^\prime} \, ,
\end{split}
 \label{eq:fleur_int_mmn_soc}
\end{equation}
where $\Theta_{\vn G}$ has been defined in Eq.~\eqref{eq:FT_theta_INT}. Compared to the implementation of the usual overlaps $M_{mn}^{(\vn k,\vn b)}$, Ref.~\onlinecite{Freimuth2008}, only reciprocal lattice vectors $\vn G$ and $\vn G^\prime$ enter $\Theta_{\vn G}$ above. Thus, we can arrive at the shape of the above overlaps by formally setting $\vn G(\vn q+\vn b)$ to zero in Eq.~\eqref{eq:fleur_int_mmn}.

In contrast to Eq.~\eqref{eq:fleur_mt_psi}, the coefficients of the expansion of the muffin tin wave functions depend on $\theta$. Accordingly, the Bloch state in the local spin-coordinate frame is given as
\begin{equation}
\begin{split}
 \Psi^\sigma_{\vn k \theta n} (\vn r)\Big|_{\text{MT}_\mu}=\sum\limits_{L}\big[ &a^{\mu,\sigma}_{Ln}(\vn k,\theta) u^{\mu,\sigma}_l(r_{\mu}) \\
+ &b^{\mu,\sigma}_{Ln}(\vn k, \theta)\dot{u}^{\mu,\sigma}_l(r_{\mu})\big] Y_L(\hat{\vn r}_{\mu}) \, ,
\end{split}
 \label{eq:fleur_mt_psi_soc}
\end{equation}
where $L$ stands for the set of angular momentum quantum numbers $(l,l_z)$. The overlaps between the lattice periodic parts, Eq.~\eqref{eq:chain_temp_beta}, are evaluated using the orthogonality of the spherical harmonics to yield
\begin{equation}
\begin{split}
 \langle u^{\sigma}_{\vn k \theta m} |& u^{\sigma^\prime}_{\vn k\, \theta+b\, n}\rangle\Big|_{\text{MT}_\mu}= \\
&=\sum\limits_{L} \left[ \left(a^{\mu,\sigma}_{Lm}(\vn k,\theta)\right)^* a^{\mu,\sigma^\prime}_{L n}(\vn k,\left[\theta+b\right]) t^{\mu,L}_{11}(\sigma,\sigma^\prime)\right. \\
 &+\left. \left(a^{\mu,\sigma}_{Lm}(\vn k,\theta)\right)^* b^{\mu,\sigma^\prime}_{L n}(\vn k,\left[\theta+b\right]) t^{\mu,L}_{12}(\sigma,\sigma^\prime)\right.\\
 &+\left.\left(b^{\mu,\sigma}_{Lm}(\vn k,\theta)\right)^* a^{\mu,\sigma^\prime}_{L n}(\vn k,\left[\theta+b\right]) t^{\mu,L}_{21}(\sigma,\sigma^\prime)\right. \\
&+\left.\left(b^{\mu,\sigma}_{Lm}(\vn k,\theta)\right)^* b^{\mu,\sigma^\prime}_{L n}(\vn k,\left[\theta+b\right]) t^{\mu,L}_{22}(\sigma,\sigma^\prime)\right] \, ,
 \end{split}
 \label{eq:fleur_sph_mmn_soc}
\end{equation}
where the coefficients $t_{ij}$ represent integrals of the radial solutions and their energy derivatives:
\begin{align}
 t^{\mu,L}_{11}(\sigma,\sigma^\prime) &= \int r_{\mu}^2 \, u^{\mu,\sigma}_l(r_{\mu}) u^{\mu,\sigma^\prime}_{l}(r_{\mu})\,\D r_{\mu}\, , \\
 t^{\mu,L}_{12}(\sigma,\sigma^\prime) &= \int r_{\mu}^2 \, u^{\mu,\sigma}_{l}(r_{\mu}) \dot{u}^{\mu,\sigma^\prime}_l(r_{\mu})\,\D r_{\mu} \, ,
\end{align}
and likewise for $t_{21}$ and $t_{22}$. Compared to Appendix~\ref{app:details_FLAPW} or the implementation of the usual overlaps $M_{mn}^{(\vn k,\vn b)}$, Ref.~\onlinecite{Freimuth2008}, the above $t$-integrals are simplified as they do not contain the Gaunt coefficients. Formally, we can obtain, for example, $t^{\mu,L}_{11}(\sigma,\sigma)$ from Eq.~\eqref{eq:t_integral_spinspiral} when setting $\vn b$ to zero.

If we consider the application to the one-dimensional magnetic chain discussed in the main text, an additional contribution arises due to the presence of the vacuum (cf. Appendix~\ref{app:od}). The wave function is expanded in the vacuum region as
\begin{equation}
 \Psi^\sigma_{k_z \theta n}(\vn r) = \sum\limits_{P} \psi^{n,\sigma}_{P}(k_z, \theta,r) \E{\I p\phi} \E{\I\left(k_z+G_z\right)z} \, ,
\label{eq:fleur_od_psi_soc}
\end{equation}
with $P=(G_z,p)$ representing the set of the integer $p$ and the plane-wave vector $G_z$, and further
\begin{equation}
\begin{split}
 \psi^{n,\sigma}_{P}(k_z, \theta, r) =& a^{n,\sigma}_{P}(k_z, \theta) u^\sigma_{P}(k_z, \theta,r) \\
+& b^{n,\sigma}_{P}(k_z, \theta)\dot{u}^{\sigma}_{P}(k_z, \theta,r) \, .
\end{split}
\end{equation}
Here, $u_P$ and $\dot{u}_P$ refer to the radial solutions of the Schr\"odinger equation in the vacuum region and their energy derivatives, respectively. Consequently, the vacuum contribution to overlaps of the periodic parts in the local frame, Eq.~\eqref{eq:chain_temp_beta}, assumes the form 
\begin{equation}
\begin{split}
 &\langle u^{\sigma}_{\vn k \theta m} | u^{\sigma^\prime}_{\vn k\, \theta+b\, n}\rangle\Big|_{\text{OD}} =\\
&=\ell\sum\limits_{P} \int_{R_\text{vac}}^{\infty} r \left(\psi^{m,\sigma}_{P}(k_z, \theta, r) \right)^* \psi^{n,\sigma^\prime}_{P}(k_z, \left[\theta+b\right], r) \, \D r \, ,
\end{split}
\end{equation}
where $\ell=2\pi T$ with $T$ as lattice constant measured along the $z$ direction, and $R_{\text{vac}}$ is the radius of the one-dimensional geometry under consideration. Unlike the case of the usual overlaps $M_{mn}^{(\vn k,\vn b)}$, Ref.~\onlinecite{Freimuth2008}, no cylindrical Bessel function occurs in the above radial integrals. The formal shape of such overlaps can therefore be obtained by considering $\vn G(q_z+\vn b)=\vn 0$ in Eq.~\eqref{eq:c6}.

\section{Derivatives of the multi-parameter Hamiltonian with respect to the additional parameter \texorpdfstring{$\gpara$}{}}
\label{app:derivative_H}
The Wannier interpolation scheme provides an elegant means of performing analytically crystal momentum derivatives of the Hamiltonian, which enter the calculation of properties such as the AHE or other transport coefficients \cite{Usui2009,Shelley2011,Wang2006,Yates2007}. We are able to compute analogously derivatives of the multi-parameter Hamiltonian $H^{(\vn k,\gpara)}$ with respect to an additional external parameter $\gpara$, starting from Eq.~\eqref{eq:ham_interpol} of the generalized Wannier interpolation:
\begin{equation}
 \frac{\partial H}{\partial \lambda_\alpha} = \sum\limits_{\vn R \gdlatt} \I\Xi_\alpha \E{\I \vn k \cdot \vn R}\E{\I \gpara \cdot \gdlatt} H(\vn R,\gdlatt) \, ,
\label{eq:H_1st_derivative}
\end{equation}
and
\begin{equation}
 \frac{\partial^2 H}{\partial \lambda_\alpha \partial \lambda_\beta} = -\sum\limits_{\vn R \gdlatt} \Xi_\alpha\Xi_\beta \E{\I \vn k \cdot \vn R}\E{\I \gpara \cdot \gdlatt} H(\vn R,\gdlatt) \, ,
\label{eq:H_2nd_derivative}
\end{equation}
where $H(\vn R,\gdlatt)$ is the matrix of the hopping elements $H_{nm}(\vn R,\gdlatt)$ between \gwfshort{}, and $\lambda_\alpha$ and $\Xi_\alpha$ refer to the $\alpha$-th components of the vectors $\gpara$ and $\gdlatt$, respectively. To simplify notation, we suppress the explicit dependence of $H^{(\vn k,\gpara)}$ on $\vn k$ and $\gpara$ here and in the following. The above equations may be particularly fruitful in accessing accurately Berry connections and curvatures.

We employ such expressions to determine the first and second derivatives of the energy $E(\gpara)$ with respect to the external parameter $\gpara$. Based on the Fermi-Dirac distribution function $f(y)$ with $y=E_F(\gpara) - \mathcal{E}_{\vn k \gpara n}$, the energy of the system is defined by
\begin{equation}
 E(\gpara)=\frac{1}{N_{\vn k}}\sum\limits_{\vn k n} \mathcal{E}_{\vn k \gpara n} f(y) \, ,
\end{equation}
with the Fermi energy $E_F(\gpara)$, and it follows that
\begin{equation}
 \partial_\alpha E(\gpara)=\frac{1}{N_{\vn k}}\sum\limits_{\vn k n} \left[\partial_\alpha \mathcal{E}_{\vn k \gpara n} f(y) + \mathcal{E}_{\vn k \gpara n} \partial_\alpha f(y)\right] \, ,
\label{eq:etot_1st_derivative}
\end{equation}
and
\begin{equation}
\begin{split}
 \partial_\alpha\partial_\beta E(\gpara)=\frac{1}{N_{\vn k}}\sum\limits_{\vn kn}[ \partial_\alpha\partial_\beta \mathcal{E}_{\vn k \gpara n} f(y) &+ \mathcal{E}_{\vn k \gpara n} \partial_\alpha\partial_\beta f(y) \\
 + \partial_\alpha \mathcal{E}_{kqn} \partial_\beta f(y) &+ \partial_\beta \mathcal{E}_{kqn} \partial_\alpha f(y) ] \, ,
\end{split}
\label{eq:etot_2nd_derivative}
\end{equation}
where the notation $\partial_\alpha = \partial/\partial \lambda_\alpha$ was introduced. We can obtain the derivatives of the band energies, which enter these equations, by using Eq.~\eqref{eq:H_1st_derivative} and Eq.~\eqref{eq:H_2nd_derivative}:
\begin{equation}
 \partial_\alpha \mathcal{E}_{\vn k \gpara n} = \left\langle \varphi_{\vn k \gpara n} | \partial_\alpha H | \varphi_{\vn k \gpara n} \right\rangle \, ,
\label{eq:e6}
\end{equation}
and
\begin{equation}
\begin{split}
 &\partial_\alpha\partial_\beta \mathcal{E}_{\vn k \gpara n} = \left\langle \varphi_{\vn k \gpara n} | \partial_\alpha\partial_\beta H| \varphi_{\vn k \gpara n} \right\rangle \\
&\quad+ 2 \Re \sum\limits_{m\neq n} \frac{\left\langle \varphi_{\vn k \gpara n} | \partial_\alpha H | \varphi_{\vn k \gpara m} \right\rangle \left\langle \varphi_{\vn k \gpara m} | \partial_\beta H| \varphi_{\vn k \gpara n} \right\rangle }{\mathcal{E}_{\vn k \gpara n}-\mathcal{E}_{\vn k \gpara m}} \, ,
\end{split}
\label{eq:e7}
\end{equation}
where the second contribution can be derived from first order pertubation theory. The states $|\varphi_{\vn k \gpara n}\rangle$ are the eigenvectors of the multi-parameter Hamiltonian $H^{(\vn k,\gpara)}$. Evaluating $\partial_\alpha f$ and $\partial_\alpha\partial_\beta f$ in Eq.~\eqref{eq:etot_1st_derivative} and Eq.~\eqref{eq:etot_2nd_derivative} requires knowledge of the derivatives of the Fermi energy $E_F(\gpara)$. To obtain analytically the necessary information, we invoke the total number of electrons in the system, $N(\gpara) = N_{\vn k}^{-1}\sum_{\vn k n} f(y)$, which is a constant, i.e., $\partial_\alpha N(\gpara)=0$. First derivatives of the Fermi energy are accordingly given by
\begin{equation}
 \partial_\alpha E_F(\gpara) = \left[\sum\limits_{\vn k n} \frac{\partial f(y)}{\partial y}\right]^{-1} \sum\limits_{\vn kn} \frac{\partial f(y)}{\partial y} \partial_\alpha \mathcal{E}_{\vn k \gpara n} \, ,
\end{equation}
where the term $\sum_{\vn k n} \partial f(y)/\partial y$ is a measure for the density of states at the Fermi level. The second derivatives of the Fermi energy assume the form
\begin{equation}
\begin{split}
&\partial_\alpha\partial_\beta E_F(\gpara) = \left[ \sum\limits_{\vn k n} \frac{\partial f(y)}{\partial y} \right]^{-1} \sum\limits_{\vn k n} \left[ \frac{\partial f(y)}{\partial y}\partial_\alpha\partial_\beta \mathcal{E}_{\vn k \gpara n} \right.\\
&\left.- \frac{\partial^2 f(y)}{\partial y^2} ( \partial_\alpha E_F(\gpara) - \partial_\alpha \mathcal{E}_{\vn k \gpara n})( \partial_\beta E_F(\gpara) - \partial_\beta \mathcal{E}_{\vn k \gpara n}) \right] \, ,
\end{split} 
\end{equation}
which is easily found from the condition $\partial_\alpha \partial_\beta N(\gpara) =0$.

At zero temperature, Eq.~\eqref{eq:etot_1st_derivative} and Eq.~\eqref{eq:etot_2nd_derivative} simplify. From the condition $\partial_\alpha N(\gpara)=0$ follows that
\begin{equation}
\begin{split}
 \partial_\alpha E(\gpara)&=\frac{1}{N_{\vn k}}\sum\limits_{\vn kn} f(y)\partial_\alpha \mathcal{E}_{\vn k \gpara n} \\
 &=\frac{1}{N_{\vn k}}\sum\limits_{\vn k n}\Theta(y) \left\langle \varphi_{\vn k \gpara n} | \partial_\alpha H | \varphi_{\vn k \gpara n} \right\rangle \, ,
\end{split}
\end{equation}
with Heaviside step function $\Theta(y)$. Likewise, we obtain
\begin{equation}
\begin{split}
 &\partial_\alpha\partial_\beta E(\gpara) =\frac{1}{N_{\vn k}}\sum\limits_{\vn k n}\left( f(y)\partial_\alpha\partial_\beta \mathcal{E}_{\vn k \gpara n} + \partial_\alpha f(y) \partial_\beta \mathcal{E}_{\vn k \gpara n} \right) \\
&=\frac{1}{N_{\vn k}}\sum\limits_{\vn k n} \Theta(y)\bigg( \left\langle \varphi_{\vn k \gpara n} | \partial_\alpha\partial_\beta H| \varphi_{\vn k \gpara n} \right\rangle \\
&\qquad\  + 2 \Re \sum\limits_{m\neq n} \frac{\left\langle \varphi_{\vn k \gpara n} | \partial_\alpha H | \varphi_{\vn k \gpara m} \right\rangle \left\langle \varphi_{\vn k \gpara m} | \partial_\beta H| \varphi_{\vn k \gpara n} \right\rangle }{\mathcal{E}_{\vn k \gpara n}-\mathcal{E}_{\vn k \gpara m}}\bigg) \\
&+\frac{1}{N_{\vn k}}\sum\limits_{\vn k n}\delta(y) \left(\partial_\alpha E_F(\gpara)-\partial_\alpha \mathcal{E}_{\vn k \gpara n}\right) \langle \varphi_{\vn k \gpara n} | \partial_\beta H | \varphi_{\vn k \gpara n}\rangle \, .
\end{split}
\label{eq:e9}
\end{equation}

To calculate accurately the derivatives of the energy $E(\gpara)$ given by Eq.~\eqref{eq:etot_1st_derivative} and Eq.~\eqref{eq:etot_2nd_derivative}, we implement the above scheme based on the hoppings. We are able to derive from generalized Wannier interpolation basic properties of the system, for example, the spin stiffness or the anisotropy constant.

\bibliography{my_bibliography}

\end{document}